\RequirePackage[l2tabu, orthodox]{nag}
\RequirePackage{snapshot}

\documentclass[10pt,onecolumn]{extarticle}

\makeatletter
\if@twocolumn
  \usepackage[dvips,letterpaper,top=0.5in, bottom=0.5in, left=0.75in, right=0.5in,includefoot,heightrounded]{geometry}
\else
  \usepackage[dvips,letterpaper,margin=1in,includefoot,heightrounded]{geometry}
\fi

\usepackage{srcltx}

\usepackage{amsmath}
\usepackage{amssymb,amsfonts}

\usepackage{abstract}

\usepackage{graphicx}
\usepackage{epsfig}
\usepackage[usenames,dvipsnames]{color}
\usepackage[usenames,dvipsnames]{xcolor}
\usepackage{subfigure}

\usepackage{booktabs}

\usepackage{setspace}
\usepackage{flushend}
\usepackage{multicol}

\usepackage{cite}
\usepackage{url}
\usepackage[normalem]{ulem}

\usepackage{enumerate}

\usepackage{hyperref}

\usepackage{lipsum}

\usepackage[sc,tiny,center]{titlesec}

\usepackage{microtype}

\makeatletter

\newcommand{\printtitle}{%
\makeatletter
\if@twocolumn

\twocolumn[%
  \maketitle
  \begin{onecolabstract}
    \myabstract
  \end{onecolabstract}
  \begin{center}
    \small
    \textbf{Keywords}
    \\\medskip
    \mykeywords
  \end{center}
  \bigskip
]
\saythanks
\else
  \maketitle
  \begin{onecolabstract}
    \myabstract
  \end{onecolabstract}
  \begin{center}
    \small
    \textbf{Keywords}
    \\\medskip
    \mykeywords
  \end{center}
  \bigskip
  \onehalfspacing
\fi
\makeatother
}

\title{%
A Digital Hardware Fast Algorithm and FPGA-based Prototype for a Novel 16-point Approximate DCT
for Image Compression Applications
}

\author{%
F.~M.~Bayer%
\thanks{%
F.~M.~Bayer
is with
the
Departamento de Estat\'{\i}stica,
Universidade Federal de Santa Maria.
E-mail:
\protect\url{bayer@ufsm.br}
}
\and
R.~J.~Cintra%
\thanks{%
R.~J.~Cintra
is with
the
Signal Processing Group,
Departamento de Estat\'{\i}stica,
Universidade Federal de Pernambuco.
E-mail:
\protect\url{rjdsc@de.ufpe.br}
}
\and
A.~Edirisuriya%
\thanks{%
A.~Edirisuriya
and
A. Madanayake
are with the
ECE, The University of Akron, Akron, OH, USA
}
\and
A.~Madanayake$^\ddagger$
}

\date{}

\newcommand{\myabstract}{%
The discrete cosine transform (DCT)
is the key step in many image and video coding standards.
The 8-point DCT is an important special case,
possessing several low-complexity approximations widely investigated.
However, 16-point DCT transform
has energy compaction advantages.
In this sense,
this paper presents
a new 16-point DCT approximation
with null multiplicative complexity.
The proposed transform matrix
is orthogonal and contains only zeros and ones.
The proposed transform
outperforms the well-know
Walsh-Hadamard transform
and the current state-of-the-art 16-point approximation.
A fast algorithm for the proposed transform is also introduced.
This fast algorithm
is experimentally validated using hardware implementations
that are physically realized and verified on a 40~nm CMOS Xilinx Virtex-6 XC6VLX240T
FPGA chip  for a maximum clock rate of 342~MHz.
Rapid prototypes on FPGA for 8-bit input word size shows significant improvement in compressed image quality by up to 1-2 dB
at the cost of only eight adders compared to the state-of-art 16-point DCT approximation algorithm in the literature
[S.~Bouguezel, M.~O. Ahmad, and M.~N.~S. Swamy.
A novel transform for image compression.
In {\em Proceedings of the 53rd IEEE International Midwest Symposium
  on Circuits and Systems (MWSCAS)}, 2010].
}

\newcommand{\mykeywords}{%
DCT Approximation,
Fast algorithms,
FPGA
}

\begin{document}

\printtitle

\section{Introduction}

The discrete cosine transform
(DCT)~\cite{ahmed1974,rao1990discrete,britanak2007discrete}
is a pivotal tool in digital signal processing,
whose
popularity is mainly due to
its good energy compaction properties.
In fact, the DCT is a robust approximation for
the optimal Karhunen-Lo\`eve transform
when first-order Markov signals, such as images,
are
considered~\cite{rao1990discrete, britanak2007discrete, liang2001}.
Indeed, the DCT has found application in several
image and video coding schemes~\cite{bhaskaran1997, britanak2007discrete},
such as
JPEG~\cite{penn1992},
MPEG-1~\cite{roma2007hybrid},
MPEG-2~\cite{mpeg2},
H.261~\cite{h261},
H.263~\cite{h263},
and
H.264~\cite{h264, h2642003, wiegand2003}.

Through the decades
signal processing literature
has been populated with
efficient methods for the DCT computation,
collectively known as fast algorithms.
This can be observed in several works
with efficient hardware and software implementations,
including~\cite{chen1977fast,suehiro1986fast,hou1087fast,arai1988fast,loeffler1989practical,Cho1991,dimitrov2004dctfree, arjuna2011algebraic}.
Methods such as the Arai DCT algorithm~\cite{arai1988fast}
can greatly reduced the number of arithmetic operations
required for the DCT evaluation.
Indeed,
current algorithms for the \emph{exact} DCT
are mature and
further
complexity reductions are very difficult to achieve.
Nevertheless,
demands for real-time video processing and transmission
are increasing~\cite{saponara2012, kuo2011}.
Therefore,
complexity reductions for the DCT must be obtained
using different methods.

One possibility is the development of approximate DCT algorithms.
Approximate transforms aim at demanding
very low complexity
while offering a close estimate of the exact calculation.
In general,
the elements of approximate transform matrices
require only $\{0, \pm 1/2, \pm 1, \pm 2 \}$~\cite{cintra2011dct}.
This implies null multiplicative complexity;
only addition and bit shifting operations are usually required.
While not computing the DCT exactly,
such approximations can provide meaningful estimations
at low-complexity requirements.

In particular,
8-point DCT approximations
have been attracting signal processing community attention.
This particular blocklength is widely adopted
in several image and video coding standards,
such as JPEG and MPEG family~\cite{penn1992, bhaskaran1997, liang2001}.
Prominent 8-point DCT approximations include
the signed discrete cosine~\cite{haweel2001new},
the
level 1 approximation by Lengwehasatit-Ortega~\cite{lengwehasatit2004scalable},
the
Bouguezel-Ahmad-Swamy (BAS) series of algorithms~\cite{bas2008,bas2009,bas2010,bas2011},
and the
DCT round-off approximations~\cite{cb2010, cintra2011dct}.
However,
transforms with blocklength greater than eight
has several advantages such as better energy compaction
and reduced quantization error~\cite{davies2010suggestion}.

In~\cite{davies2010suggestion},
an adapted version of
the 16-point
Chen's fast DCT algorithm~\cite{chen1977fast,rao1990discrete}
is suggested for video encoding.
Chen's algorithm requires multiplicative constants
$\cos(k\pi/32)$, $k=1,2,\ldots,15$,
which can be approximated by
fixed precision quantities~\cite[Sec.~5]{davies2010suggestion}.
Indeed,
dyadic rational were employed~\cite{britanak2007discrete},
resulting in a non-orthogonal
transform~\cite[Sec.~5]{davies2010suggestion}.
The International Telecommunication Union
fosters image blocks of 16$\times$16
pixels~\cite{telecommunicationstandardizationsector2009video}
instead of the 4$\times$4 and 8$\times$8 pixel blocks
required by the H.264/MPEG-4 AVC
standard for video compression~\cite{malvar2003low-complexity}.
The main reason for such recommendation
is the improved coding gains~\cite{lee2008technical}.
It is clear that for such large transform blocklengths,
minimizing the computational complexity
becomes
a central issue~\cite{davies2010suggestion}.

In this context,
the main goal of this paper is to advance
16-point approximate DCT architectures.
First,
we introduce a new low-complexity 16-point DCT approximation.
The proposed transform is sought to be orthogonal
and to possess null multiplicative complexity.
Second,
we propose an efficient fast algorithm for the new transform.
Third,
we introduce hardware implementations for
the proposed transform as well as for the
16-point DCT approximate method
introduced by Bouguezel-Ahmad-Swamy (BAS-2010) in~\cite{bas2010}.
Both methods
are demonstrated to be
suitable for image compression.

The paper unfolds as follows.
In Section~\ref{s:app},
the new proposed transform is introduced and
mathematically analyzed.
Error metrics are considered to assess its proximity
to the exact DCT matrix.
In Section~\ref{section-fast-algorithm},
a fast algorithm for the proposed transform is
derived and its computational complexity is compared with
existing methods.
An image compression simulation is described in
Section~\ref{s:compression},
indicating the adequateness of the introduced transform.
In Section~\ref{section-implementation},
FPGA-based hardware implementations for
both the proposed transform and the
BAS-2010 approximation
are
detailed and analyzed.
Conclusions and final remarks are given in
Section~\ref{section-conclusion}.

\section{16-point DCT Approximation }
\label{s:app}

In this section,
a new 16-point multiplication-free transform is presented.
The proposed matrix transform $\mathbf{T}$
was obtained by judiciously replacing each floating point
of the 16-point DCT matrix
for 0, 1, or $-1$.
Substitutions were computationally
performed in such a way that:
(i)~the resulting matrix could satisfy
the following orthogonality-like property:
\begin{equation*}
\mathbf{T} \times \mathbf{T}^{\top}
=
\mbox{diagonal matrix},
\end{equation*}
(ii)~DCT symmetries could be preserved,
and
(iii)~the resulting matrix could offer good
energy compaction properties~\cite{haweel2001new}.
Among the several possible outcomes,
we isolated the following matrix:
\begin{equation}
\label{equation-T}
{\scriptsize
\mathbf{T}
=
\left[
\begin{array}{rrrrrrrrrrrrrrrr}
1&1&1&1&1&1&1&1&1&1&1&1&1&1&1&1\\
1&1&1&1&1&0&1&1&-1&-1&0&-1&-1&-1&-1&-1\\
1&1&1&0&0&-1&-1&-1&-1&-1&-1&0&0&1&1&1\\
1&1&1&0&-1&-1&-1&-1&1&1&1&1&0&-1&-1&-1\\
1&1&-1&-1&-1&-1&1&1&1&1&-1&-1&-1&-1&1&1\\
1&1&-1&-1&-1&1&1&0&0&-1&-1&1&1&1&-1&-1\\
1&0&-1&-1&1&1&0&-1&-1&0&1&1&-1&-1&0&1\\
1&0&-1&1&1&1&-1&-1&1&1&-1&-1&-1&1&0&-1\\
1&-1&-1&1&1&-1&-1&1&1&-1&-1&1&1&-1&-1&1\\
1&-1&-1&1&-1&-1&0&1&-1&0&1&1&-1&1&1&-1\\
1&-1&0&1&-1&0&1&-1&-1&1&0&-1&1&0&-1&1\\
0&-1&1&1&-1&1&1&-1&1&-1&-1&1&-1&-1&1&0\\
1&-1&1&-1&-1&1&-1&1&1&-1&1&-1&-1&1&-1&1\\
1&-1&1&-1&0&1&-1&1&-1&1&-1&0&1&-1&1&-1\\
0&-1&1&-1&1&-1&1&0&0&1&-1&1&-1&1&-1&0\\
1&-1&0&-1&1&-1&1&-1&1&-1&1&-1&1&0&1&-1
\end{array}
\right]
}
.
\end{equation}
Above matrix furnishes a DCT approximation given by
\begin{equation*}
\hat{\mathbf{C}}= \mathbf{D} \cdot \mathbf{T},
\end{equation*}
where
\begin{eqnarray*}
\mathbf{D} =
\mathrm{diag}
&
\left(
\frac{1}{4},
\frac{1}{\sqrt{14}},
\frac{1}{2\sqrt{3}},
\frac{1}{\sqrt{14}},
\frac{1}{4},
\frac{1}{\sqrt{14}},
\frac{1}{2\sqrt{3}},
\frac{1}{\sqrt{14}},
\right.
\\
&
\enspace
\left.
\frac{1}{4},
\frac{1}{\sqrt{14}},
\frac{1}{2\sqrt{3}},
\frac{1}{\sqrt{14}},
\frac{1}{4},
\frac{1}{\sqrt{14}},
\frac{1}{2\sqrt{3}},
\frac{1}{\sqrt{14}}
\right)
,
\end{eqnarray*}
and
$\mathrm{diag}(\cdot)$ returns the block diagonal concatenation of
its arguments.

The proposed transform $\hat{\mathbf{C}}$
is orthogonal
and requires no multiplications or bit shifting operations.
Only additions are required for
the computation of the proposed DCT approximation.
Moreover,
the scaling matrix~$\mathbf{D}$ may not introduce
any additional computational overhead in the
context of image compression.
In fact,
the scalar multiplications of~$\mathbf{D}$ can be merged into
the quantization
step~\cite{lengwehasatit2004scalable,bas2008,bas2009,bas2011,cintra2011dct}.
Therefore,
in this sense,
the approximation $\hat{\mathbf{C}}$
has the same low computational complexity of $\mathbf{T}$.

Now we aim at comparing the proposed transformation
with existing low-complexity approximations for the 16-point DCT.
Although there is a reduced number of
16-point transforms with null multiplicative complexity
in signal processing literature,
we could separate two orthogonal transformations for comparison:
(i)~the well-known Walsh-Hadamard
transform (WHT)~\cite[p.~1087]{salomon2011computer}
and
(ii)~the
16-point BAS-2010 approximation~\cite{bas2010}.
The WHT is selected for
its simplicity of implementation~\cite[p.~472]{gonzalez2002digital}.
The BAS-2010 method considered since it is the most recent method
for DCT approximation for 16-point long data.

A classical reference
in this field is the signed DCT (SDCT)~\cite{haweel2001new},
which became a standard for comparison when considering
8-point DCT approximations.
However,
for 16-point data,
the signed DCT is not orthogonal and
its inverse transformation
requires several additions and multiplications~\cite{bas2010}.
Thus, we could not consider SDCT for any meaningful comparison.

According to the
methodology employed in~\cite{haweel2001new} and
supported by~\cite{cintra2011dct},
we can assess how adequate
the proposed approximation is.
For such analysis,
each row of a
16$\times$16
approximation matrix~$\mathbf{A}$
can be interpreted as
the coefficients of a FIR filter.
Therefore,
the following filters are defined:
\begin{equation*}
h_m[n] = a_{m,n},
\quad
m=0,1,\ldots,15,
\end{equation*}
where
$a_{m,n}$ is the $(m+1,n+1)$-th entry of $\mathbf{A}$.

Thus,
the transfer functions
associated to
$h_m[n]$, $m=0,1,\ldots,15$,
can computed by the discrete-time Fourier transform
defined over $\omega \in [0,\pi]$~\cite{oppenheim2009discrete-time}:
\begin{equation*}
H_m(\omega;\mathbf{A})
=
\sum_{n=0}^{15}
h_m[n]
\exp(-j \, n \, \omega),
\quad
m=0,1,\ldots,15,
\end{equation*}
where
$j=\sqrt{-1}$.

Spectral data $H_m(\omega;\mathbf{A})$ can be employed
to define a figure of merit
for
assessing DCT approximations.
Indeed,
we can measure
the distance between
$H_m(\omega; \mathbf{C})$
and
$H_m(\omega; \mathbf{A})$,
where $\mathbf{C}$ is the \emph{exact} DCT matrix.
We adopted the
squared magnitude as a distance measure function.
Thus,
we obtained
the following mathematical expression:
\begin{equation*}
D_m(\omega;\mathbf{A})
\triangleq
\Big|
H_m(\omega;\mathbf{C})
-
H_m(\omega;\mathbf{A})
\Big|^2
,
\quad
m=0,1,\ldots,15,
\end{equation*}
Note that $D_m(\omega;\mathbf{A})$ is an energy-related error measure.
For each row $m$ at any angular frequency~$\omega \in [0,\pi]$
in radians per sample,
above expression
quantifies how discrepant
a given approximation matrix~$\mathbf{A}$ is from the DCT matrix.

\begin{figure*}%
\centering
\includegraphics[width=0.85\textwidth]{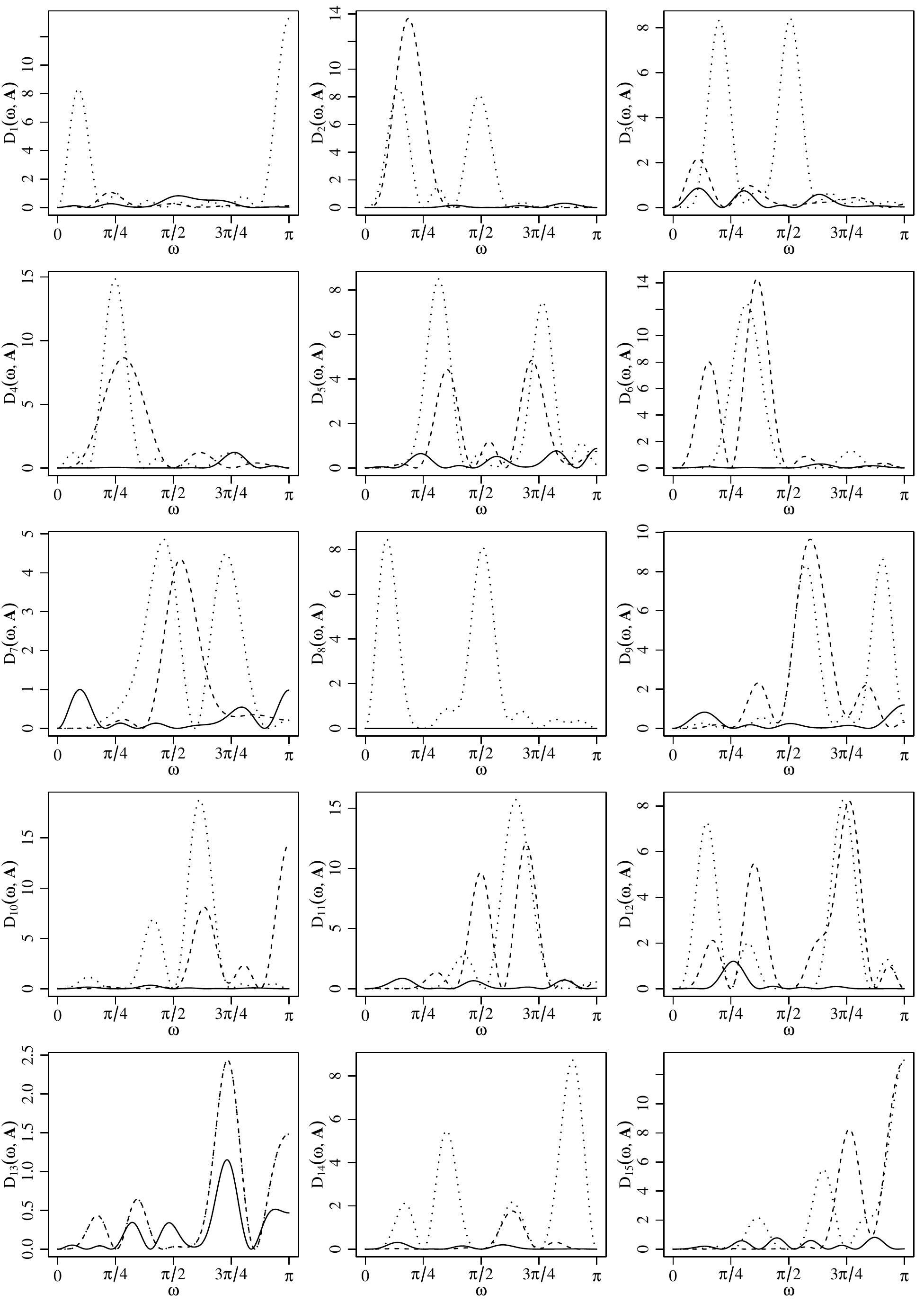}
\caption{
Plots of $D_m(\omega;\mathbf{A})$
for $m=1,2,\ldots,15$
and $\omega\in[0,\pi]$,
considering
the proposed transform (solid curve),
the BAS-2010 transform (dashed curve),
and
the WHT (dotted curve).
}
\label{f:square_magnitude_error}
\end{figure*}

Fig.~\ref{f:square_magnitude_error}
shows the plots
for
$D_m(\omega;\mathbf{A})$,
$m=1,2,\ldots,15$,
where $\mathbf{A}$ is
either
the WHT,
the BAS-2010 approximation,
or
the proposed transform~$\mathbf{T}$.
The particular plot for $m=0$ was suppressed,
since all considered transforms could match
the DCT exactly.

The
error energy
departing from the actual DCT
can be obtained by integrating
$D_m(\omega;\mathbf{A})$
over $\omega \in [0,\pi]$~\cite{cintra2011dct}:
\begin{equation*}
\label{equation-total-error-energy}
\epsilon_m(\mathbf{A})
=
\int_0^\pi
D_m(\omega;\mathbf{A})
\mathrm{d}\omega,
\end{equation*}
Table~\ref{t:error} summarizes
the obtained values of
$\epsilon_m(\mathbf{A})$,
$m=0,1,\ldots,15$.
These quantities were
computed by numerical quadrature methods~\cite{piessens1983quadpack}.

\begin{table}
\centering
\caption{Error energy $\epsilon_m(\mathbf{A})$
for selected DCT approximatinons.}
\label{t:error}
\begin{tabular}{lcccc}
\toprule
$m$ & Proposed & BAS-2010 & WHT \\
\midrule
0	&	0.00	&	0.00	&	0.00	\\
1	&	0.78	&	0.61	&	6.78	\\
2	&	0.20	&	6.34	&	6.09	\\
3	&	0.78	&	1.52	&	5.20	\\
4	&	0.50	&	6.28	&	5.70	\\
5	&	0.95	&	4.31	&	5.44	\\
6	&	0.22	&	8.60	&	6.60	\\
7	&	0.87	&	2.69	&	4.81	\\
8	&	0.00	&	0.00	&	5.50	\\
9	&	0.79	&	6.57	&	6.57	\\
10	&	0.19	&	6.69	&	10.62	\\
11	&	0.72	&	8.47	&	8.06	\\
12	&	0.46	&	5.84	&	7.29	\\
13	&	0.70	&	1.60	&	1.60	\\
14	&	0.22	&	0.75	&	5.97	\\
15	&	0.70	&	6.72	&	6.42	\\
\midrule
Total	&	8.08	&	66.99	&	92.65	\\
\toprule

\end{tabular}
\end{table}

\section{Fast Algorithm}
\label{section-fast-algorithm}

As defined in~(\ref{equation-T}),
transformation matrix~$\mathbf{T}$
requires~208 additions,
which a significant number of operations.
In the following,
we present a factorization of~$\mathbf{T}$
obtained by means of butterfly-based methods
in a decimation-in-frequency structure~\cite{blahut1985fast}.
For notational purposes,
we denote
$\mathbf{I}_n$ as the identity matrix of order~$n$,
$\bar{\mathbf{I}}_n$ as the opposite diagonal identity matrix of order~$n$,
and
the butterfly matrix as
\begin{equation*}
\mathbf{B}_n
\triangleq
\left[\begin{array}{rr}
\mathbf{I}_{n/2} & \bar{\mathbf{I}}_{n/2} \\
\bar{\mathbf{I}}_{n/2} & -\mathbf{I}_{n/2}
\end{array}
\right]
.
\end{equation*}

We maintain that
$\mathbf{T}$
can be decomposed
into less complex matrix terms
according to
the following factorization:
\begin{eqnarray*}
\mathbf{T}
=&
\mathbf{P}
\times
\mathrm{diag}
\left(
\mathbf{B}_2, \bar{\mathbf{B}}_2,
\mathbf{E},
\mathbf{O}
\right)
\times
\mathrm{diag}(\mathbf{B}_4, \mathbf{I}_{12})
\times
\mathrm{diag}(\mathbf{B}_8, \mathbf{I}_8)
\times
\mathbf{B}_{16}
,
\end{eqnarray*}
where
the
required matrices are described below:
\begin{equation*}
\bar{\mathbf{B}}_2
=
\mathbf{B}_2
\times
\bar{\mathbf{I}}_2
=
\left[
\begin{array}{rr}
1  &   1 \\
-1  &   1
\end{array}
\right]
,
\end{equation*}
\begin{equation*}
\mathbf{E}
=
\left[
\begin{array}{rrrr}
     0  &   1  &   1  &   1 \\
    -1  &  -1  &   0  &   1 \\
     1  &   0  &  -1  &   1 \\
    -1  &   1  &  -1  &   0
\end{array}
\right]
,
\end{equation*}
\begin{equation}
\label{equation-matrix-odd}
\mathbf{O}
=
\left[
\begin{array}{rrrrrrrr}
     1  &   1  &   0  &   1  &   1  &   1  &   1  &   1 \\
    -1  &  -1  &  -1  &  -1  &   0  &   1  &   1  &   1 \\
     0  &   1  &   1  &  -1  &  -1  &  -1  &   1  &   1 \\
    -1  &  -1  &   1  &   1  &   1  &  -1  &   0  &   1 \\
     1  &   0  &  -1  &  -1  &   1  &  -1  &  -1  &   1 \\
    -1  &   1  &   1  &  -1  &   1  &   1  &  -1  &   0 \\
     1  &  -1  &   1  &   0  &  -1  &   1  &  -1  &   1 \\
    -1  &   1  &  -1  &   1  &  -1  &   0  &  -1  &   1
\end{array}
\right]
,
\end{equation}
and
matrix~$\mathbf{P}$ is a permutation matrix given by
\begin{eqnarray*}
\mathbf{P}
=
&
\left[
\begin{array}{c|c|c|c|c|c|c|c|}
\mathbf{e}_1 &
\mathbf{e}_9 &
\mathbf{e}_5 &
\mathbf{e}_{13} &
\mathbf{e}_3 &
\mathbf{e}_7 &
\mathbf{e}_{11} &
\mathbf{e}_{15}
\end{array}
\right.
\\
&
\:\:\!
\left.
\begin{array}{c|c|c|c|c|c|c|c}
\mathbf{e}_2 &
\mathbf{e}_4 &
\mathbf{e}_6 &
\mathbf{e}_8 &
\mathbf{e}_{10} &
\mathbf{e}_{12} &
\mathbf{e}_{14} &
\mathbf{e}_{16}
\end{array}
\right]
,
\end{eqnarray*}
where
$\mathbf{e}_j$ is
a 16-point column vector with one in position $j$ and zero elsewhere.

Matrix $\mathbf{E}$ corresponds to the even-odd part,
whereas
matrix~$\mathbf{O}$ is linked to the odd part of the
proposed transformation~\cite[p.~71]{bi2004transforms}.
A row permuted version of matrix $\mathbf{E}$
was already reported in literature
in the derivation of the 8-point DCT approximation
described in~\cite[Fig.~1]{cintra2011dct}.

On the other hand,
matrix~$\mathbf{O}$ does not seem to be reported.
Without any further consideration,
matrix~$\mathbf{O}$ requires 48 additions.
The locations of zero elements in~(\ref{equation-matrix-odd})
is such that
a decimation-in-frequency operation by means of a butterfly structure
is prevented.
In order to obtain the required symmetry,
we propose the following manipulation:
\begin{equation}
\label{equation-odd-s}
\mathbf{O}' = \mathbf{O} - \mathbf{S}
,
\end{equation}
where
\begin{equation*}
\mathbf{S}
=
\left[
\begin{array}{rrrrrrrr}
     0  &   \phantom{-}0  &   \phantom{-}0  &   \phantom{-}1  &   0  &   \phantom{-}0  &   0  &   \phantom{-}0 \\
     0  &   0  &   0  &   1  &   0  &   0  &   0  &   0 \\
     0  &   0  &   0  &   0  &   0  &   1  &   0  &   0 \\
     0  &   1  &   0  &   0  &   0  &   0  &   0  &   0 \\
     0  &   0  &   0  &   0  &   0  &   0  &   0  &   1 \\
     1  &   0  &   0  &   0  &   0  &   0  &   0  &   0 \\
     0  &   0  &   0  &   0  &   0  &   0  &  -1  &   0 \\
     0  &   0  &   1  &   0  &   0  &   0  &   0  &   0 \\
     0  &   0  &   0  &   0  &  -1  &   0  &   0  &   0
\end{array}
\right]
.
\end{equation*}
The resulting matrix~$\mathbf{O}'$ can factorized
according to:

\begin{equation}
\label{equation-odd-prime}
\mathbf{O}'
=
\left[
\begin{array}{rrrrrrrr}
     1  &   0  &   0  &   0  &   1  &   0  &   \phantom{-}1  &   0 \\
    -1  &   0  &  -1  &   0  &   0  &   0  &   1  &   0 \\
     0  &   0  &   0  &   1  &  -1  &   0  &   1  &   0 \\
    -1  &   0  &   1  &   0  &   0  &   1  &   0  &   0 \\
     0  &   0  &  -1  &   0  &   0  &   1  &   0  &  -1 \\
     0  &  -1  &   0  &   1  &   1  &   0  &   0  &   0 \\
     0  &   1  &   0  &   0  &   0  &  -1  &   0  &  -1 \\
     0  &  -1  &   0  &  -1  &   0  &   0  &   0  &  -1
\end{array}
\right]
\times
(
\mathbf{I}_4
\otimes
\mathbf{B}_2
)
,
\end{equation}
where $\otimes$ denotes the Kronecker product.
The additive complexity of matrix~$\mathbf{O}'$ is 20~additions.

\begin{figure}
\centering
\subfigure[Full diagram]{\includegraphics{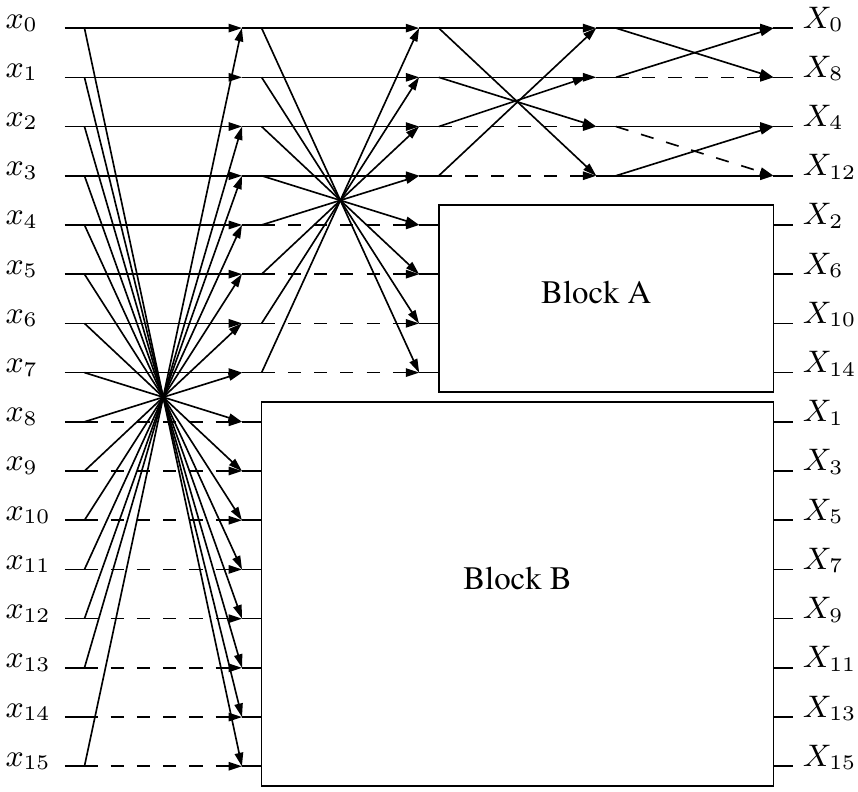}} \\
\subfigure[Block A]{\includegraphics{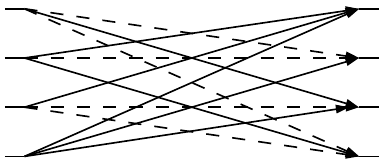}}
\caption{Flow diagram of the fast algorithm for the proposed transform.}
\label{f:fast}
\end{figure}

\begin{figure}
\centering
\subfigure[Block B]{\includegraphics{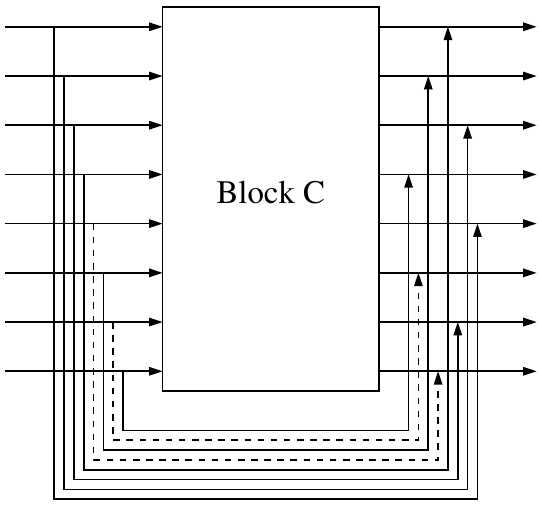}}
\qquad
\subfigure[Block C]{\includegraphics{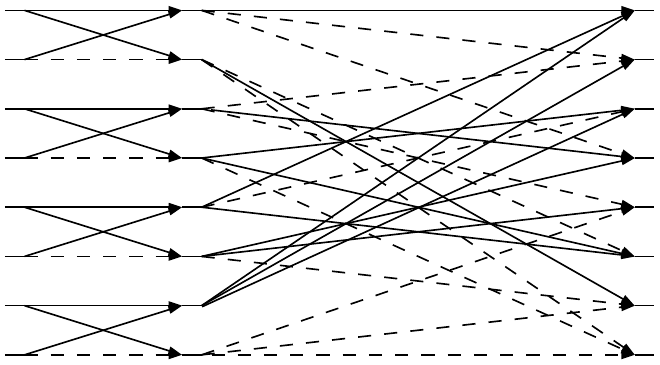}}
\caption{Flow diagram of the inner structure of Block~B.}
\label{figure-block-b}
\end{figure}

Above mathematical description can be given
a flow diagram,
which is useful for subsequent hardware implementation.
Fig.~\ref{f:fast}(a) depicts the general
structure of proposed fast algorithm.
Block~A and Block~B
represent the operations associated to
matrix $\mathbf{E}$ and $\mathbf{O}$,
respectively.
The structure of Block~A is disclosed in Fig.~\ref{f:fast}(b).
Fig.~\ref{figure-block-b}(a)
details
the inner structure of Block~B
as described in~(\ref{equation-odd-s}).
Fig.~\ref{figure-block-b}(b)
exhibits
Block~C
according to~(\ref{equation-odd-prime}).

The proposed algorithm requires only 72 additions.
Bit shifting and multiplication operations are absent.
Arithmetic complexity comparisons
with selected 16-point transforms are summarized in Table~\ref{t:complex}.

\begin{table}
\centering
\caption{Arithmetic complexity analysis.}
\label{t:complex}
\begin{tabular}{lccc}
\toprule
Operation &
Proposed &
BAS-2010~\cite{bas2010} &
WHT~\cite{fino1972relations} \\
\midrule
 Addition & 72 & 64 & 64 \\
 Bit shifting & 0 & 8 & 0  \\
 Multiplication & 0 & 0 & 0  \\
\midrule
 Total & 72 & 72 & 64  \\
\toprule
\end{tabular}
\end{table}

\section{Application to Image Compression}\label{s:compression}

This section presents the application
of the proposed transform to image compression.
We produce evidence that
it outperforms the other transforms in consideration.
For this analysis,
we used the methodology
described in~\cite{haweel2001new},
supported in~\cite{bas2008,bas2009,bas2010,bas2011},
and extended in~\cite{cintra2011dct}.

A set of 45 512$\times$512 8-bit greyscale images obtained
from a standard public image bank~\cite{uscsipi} was considered.
We adapted the JPEG compression technique~\cite{penn1992}
for the 16$\times$16 matrix case.
Each image was divided into 16$\times$16 sub-blocks,
which were submitted to the two-dimensional (\mbox{2-D})
transform
procedure associated to
the DCT matrix,
the BAS-2010~\cite{bas2010} matrix,
the WHT~\cite{fino1972relations} matrix,
and
the proposed matrix $\hat{\mathbf{C}}$.
A 16$\times$16 image block $\mathbf{K}$ has its
\mbox{2-D} transform mathematically
expressed by~\cite{suzuki2010integer}:
\begin{equation*}
\mathbf{A} \cdot \mathbf{K} \cdot \mathbf{A}^\top
,
\end{equation*}
where $\mathbf{A}$ is a considered transformation.

This computation furnished 256 approximate
transform domain coefficients for each sub-block.
A hard thresholding step was applied,
where only the $r$ initial coefficients were retained,
being the remaining ones set to zero.
Coefficients were ordered according to the usual
zig-zag scheme extended to 16$\times$16
image blocks~\cite{pao1998approximation}.
We adopted $r \in \{2,4,\ldots,254,256\}$.
The inverse procedure was then applied to reconstruct the processed data
and image quality was assessed.

Image degradation was evaluated using
three different quality measures:
(i)~the peak signal-to-noise ratio~(PSNR),
(ii)~the mean square error~(MSE),
and
(iii)~the universal quality index~(UQI)~\cite{wang2002universal}.
The PSNR and MSE were selected due to
their wide application as figures of merit in image processing.
The
UQI is considered
an improvement over PSNR and~MSE
as a tool for image quality assessment~\cite{wang2002universal}.
The UQI includes luminance, contrast, and structure characteristic
in its definition.
Another possible metric is
the structural-similarity-based image quality assessment~(SSIM)~\cite{wang2004}.
Being a variation of the UQI,
SSIM results were not very different from the measurements offered by the UQI
for the considered images.
Indeed, whenever a difference was present, it was in the order of $10^{-4}$ only.
Therefore,
SSIM results are not presented here.

Moreover,
in contrast with the JPEG image compression simulations
described in~\cite{bas2008,bas2009,bas2010,bas2011},
we considered the average measures
from all images instead of the results
derived from selected images.
In fact,
average calculations may furnish more robust
results~\cite{kay1993fundamentals}.

\begin{figure}
\centering
\subfigure[Average PSNR]
{\includegraphics[width=0.45\linewidth]{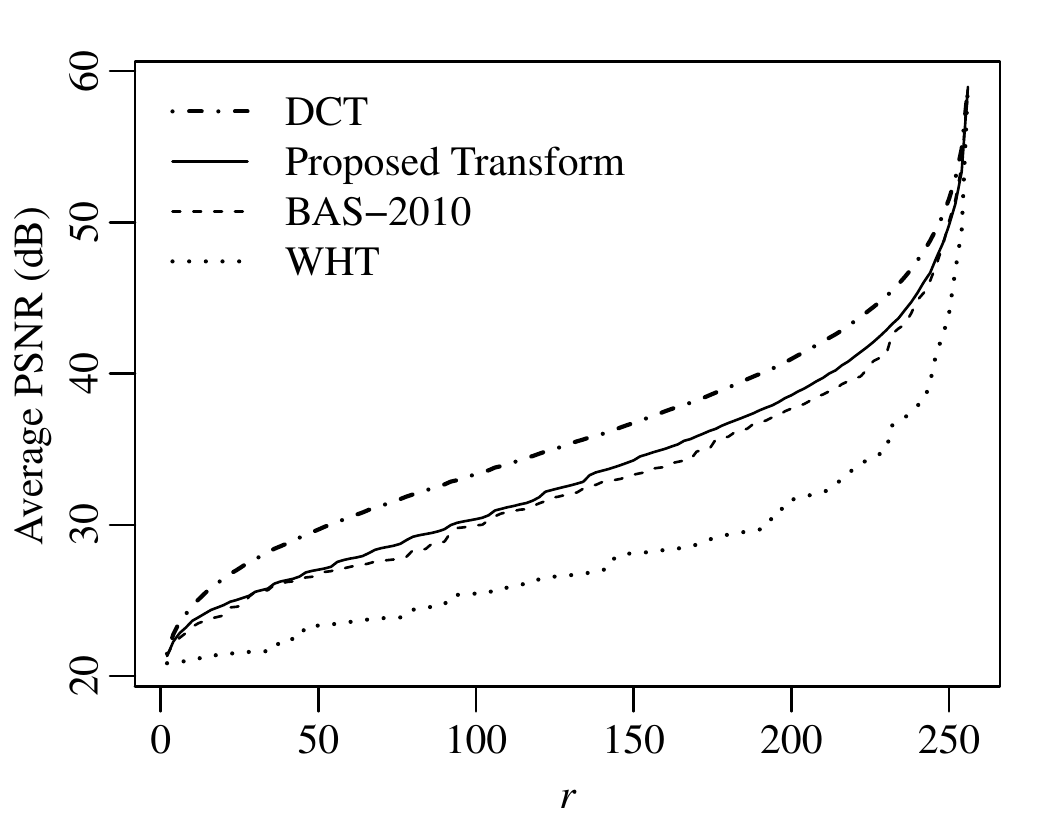} \label{psnr}}
\subfigure[PSNR difference relative to DCT (dB)]
{\includegraphics[width=0.45\linewidth]{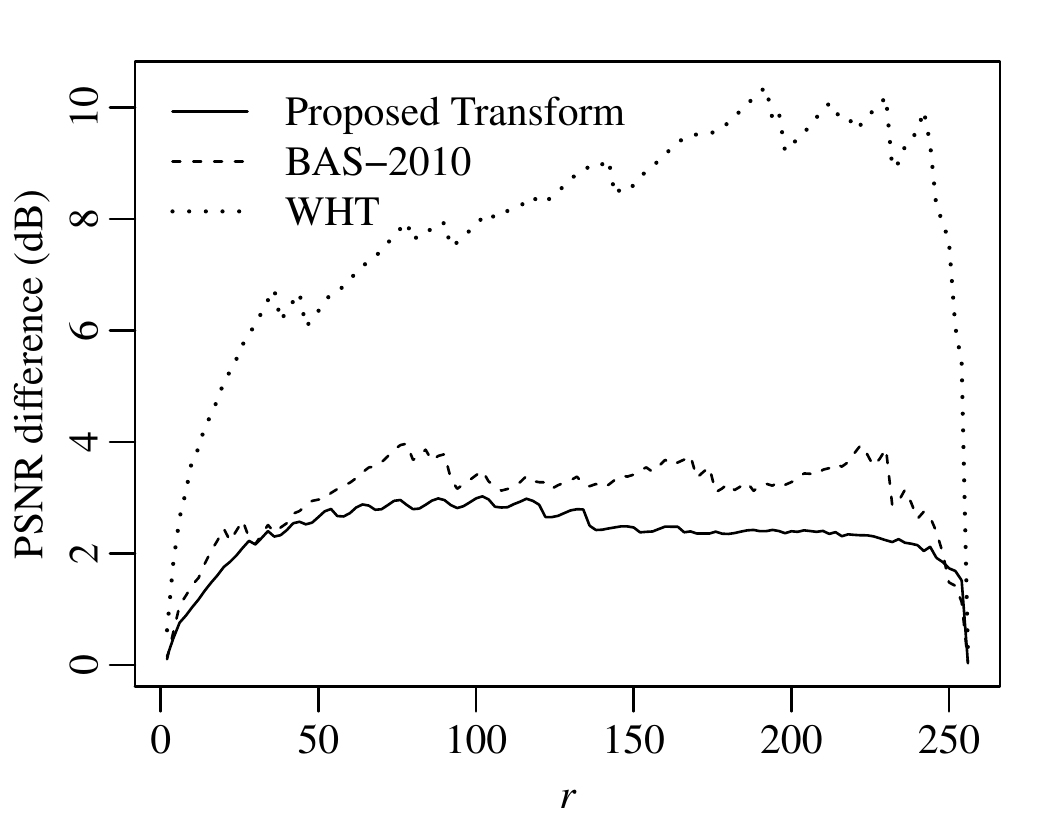} \label{psnr_diff}}
\subfigure[Average MSE]
{\includegraphics[width=0.45\linewidth]{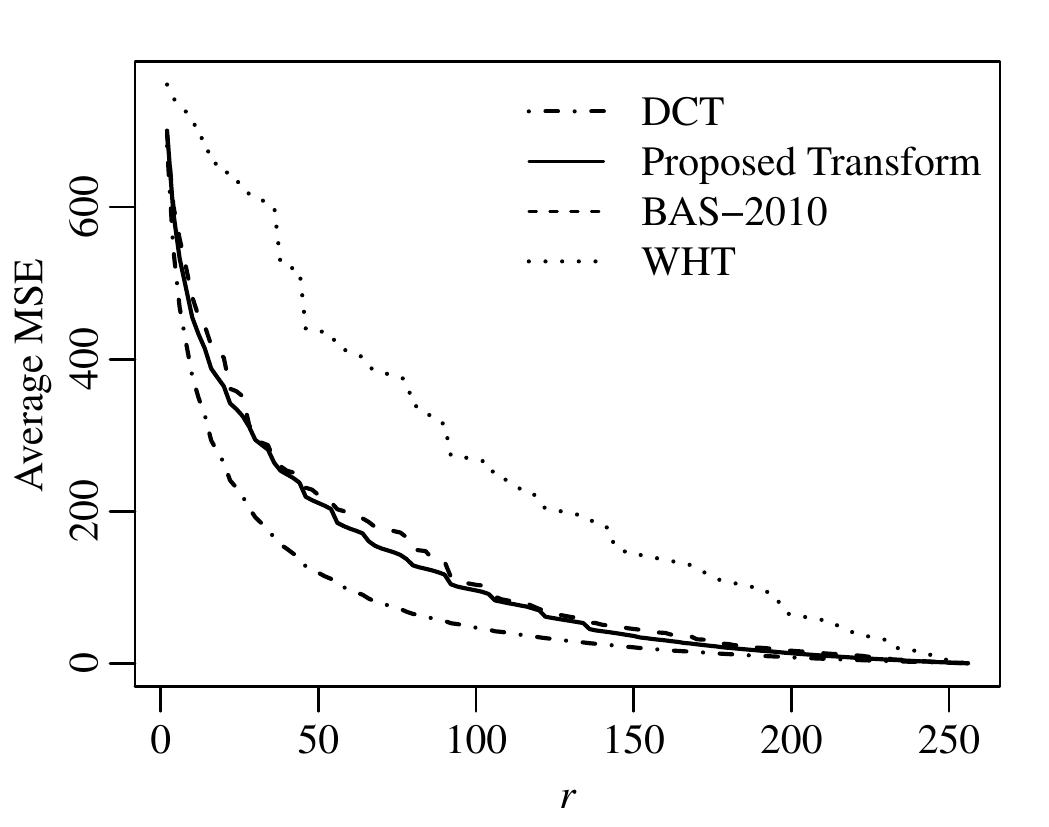} \label{mse}}
\subfigure[MSE difference relative to DCT]
{\includegraphics[width=0.45\linewidth]{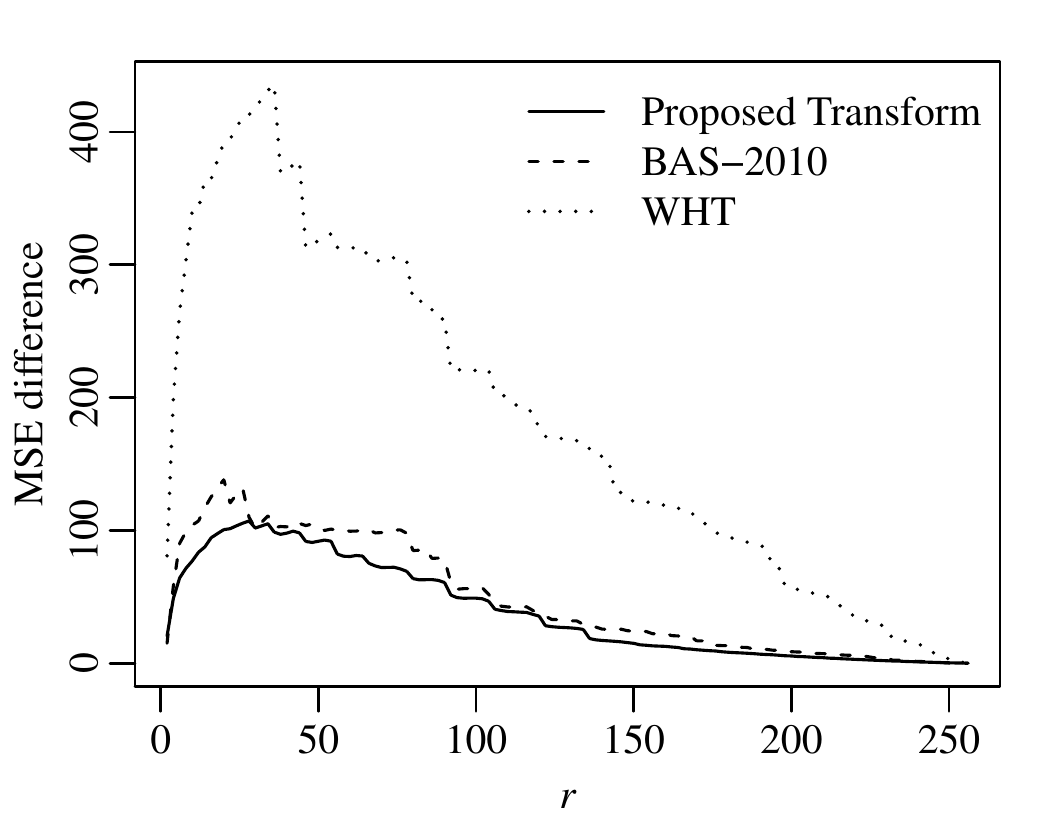} \label{mse_diff}}
\subfigure[Average UQI]
{\includegraphics[width=0.45\linewidth]{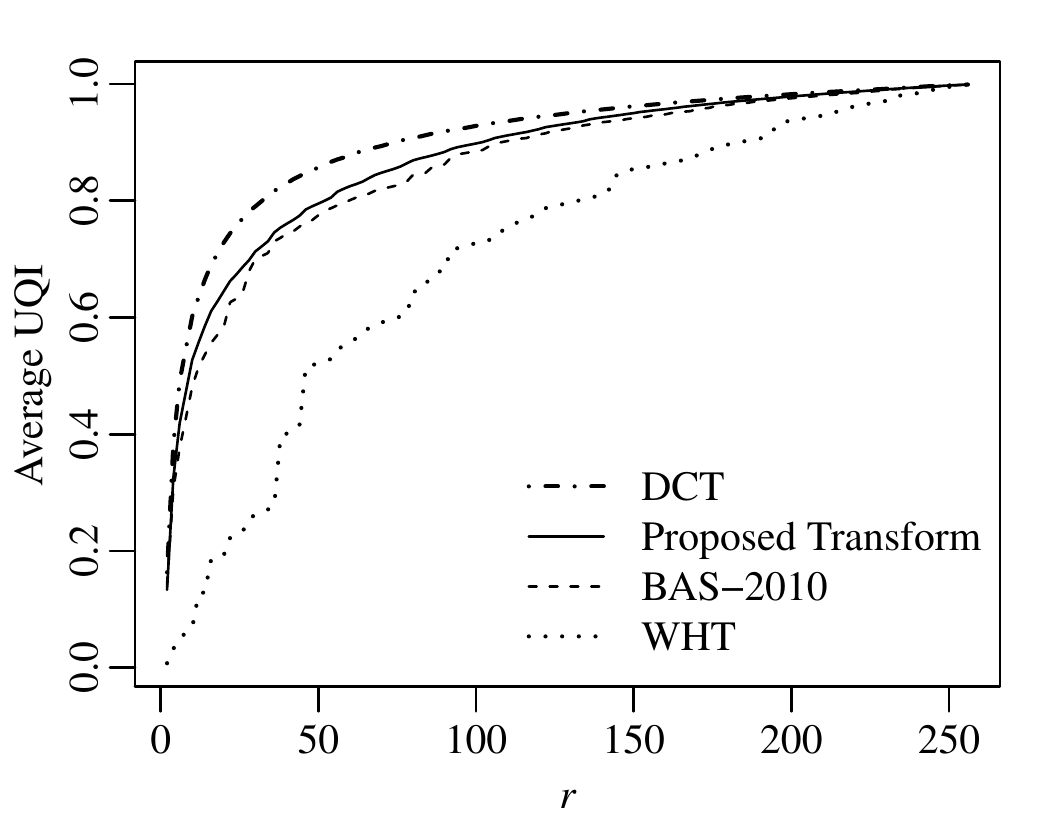} \label{uqi}}
\subfigure[UQI difference relative to DCT]
{\includegraphics[width=0.45\linewidth]{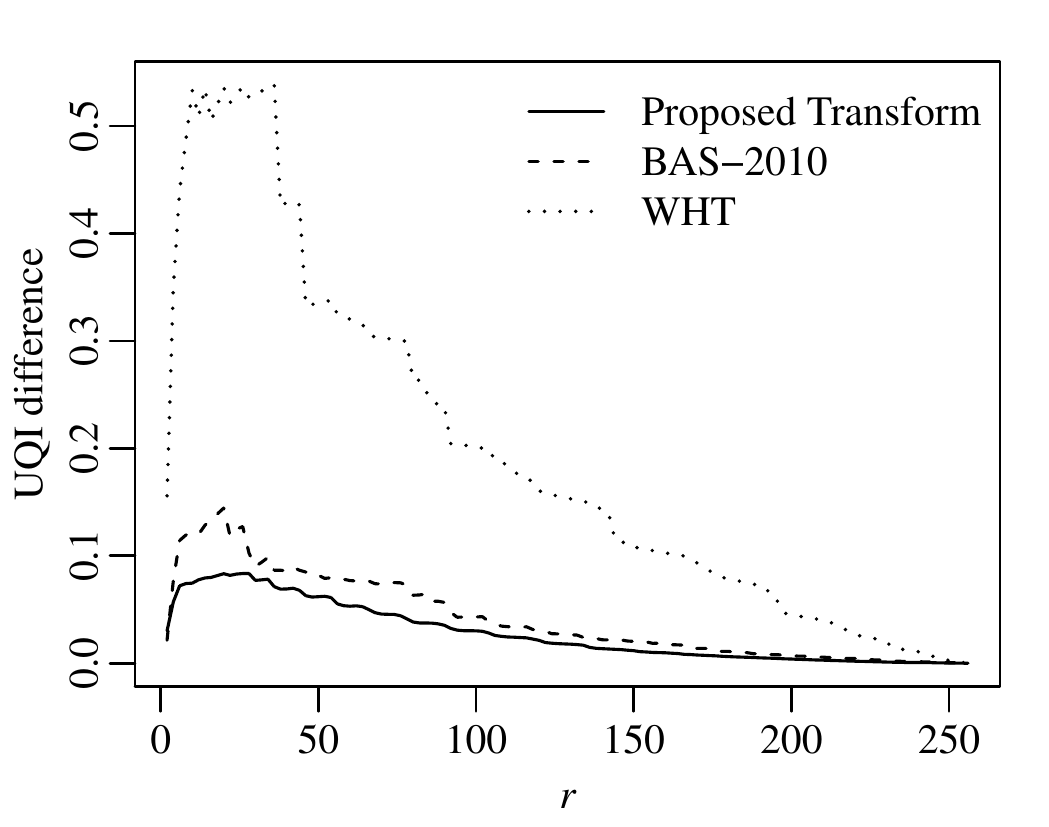} \label{uqi_diff}}
\caption{
Quality measures for several compression ratios.
}
\label{f:quality}
\end{figure}

Fig.~\ref{f:quality} shows the resulting quality measures.
The proposed transform could outperform both
the BAS-2010 transform
and
the WHT
in all compression rates
according to all considered quality measures.
Fig.~\ref{psnr_diff} shows that the proposed transform
outperformed the BAS-2010 transform in $\approx 1$~dB
and the WHT in $\approx 8$~dB,
which corresponds to $\approx 26\%$ and $\approx 630\%$
gains, respectively.
At the same time,
Fig.~\ref{psnr_diff} shows that
the results of the proposed transform
are
at most 2~dB way
from
when DCT results
at compression ratios superior to $85\%$ ($r<40$).

\begin{figure}%
\centering

\subfigure[DCT]{\includegraphics[width=0.24\linewidth]{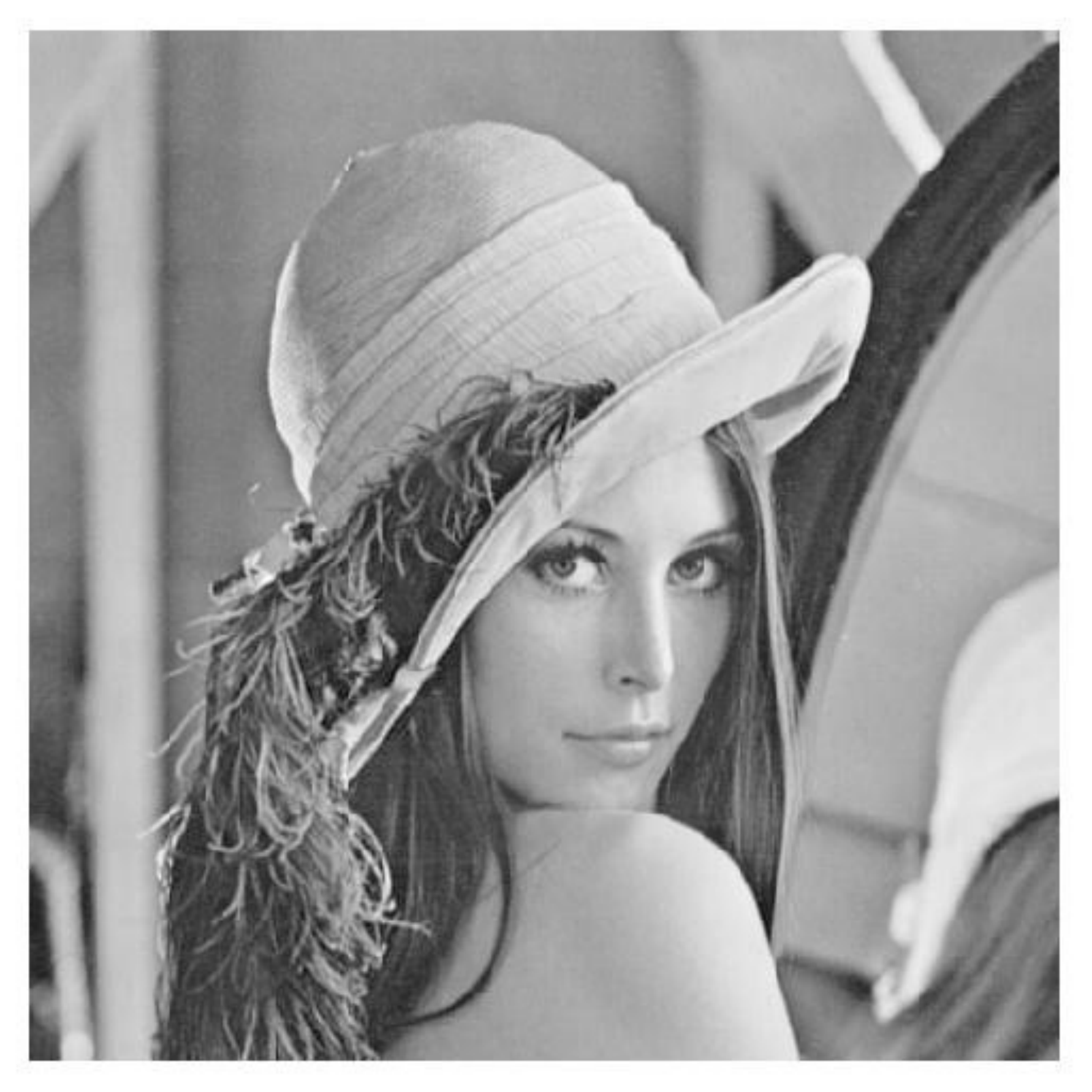}}
\subfigure[proposed]{\includegraphics[width=0.24\linewidth]{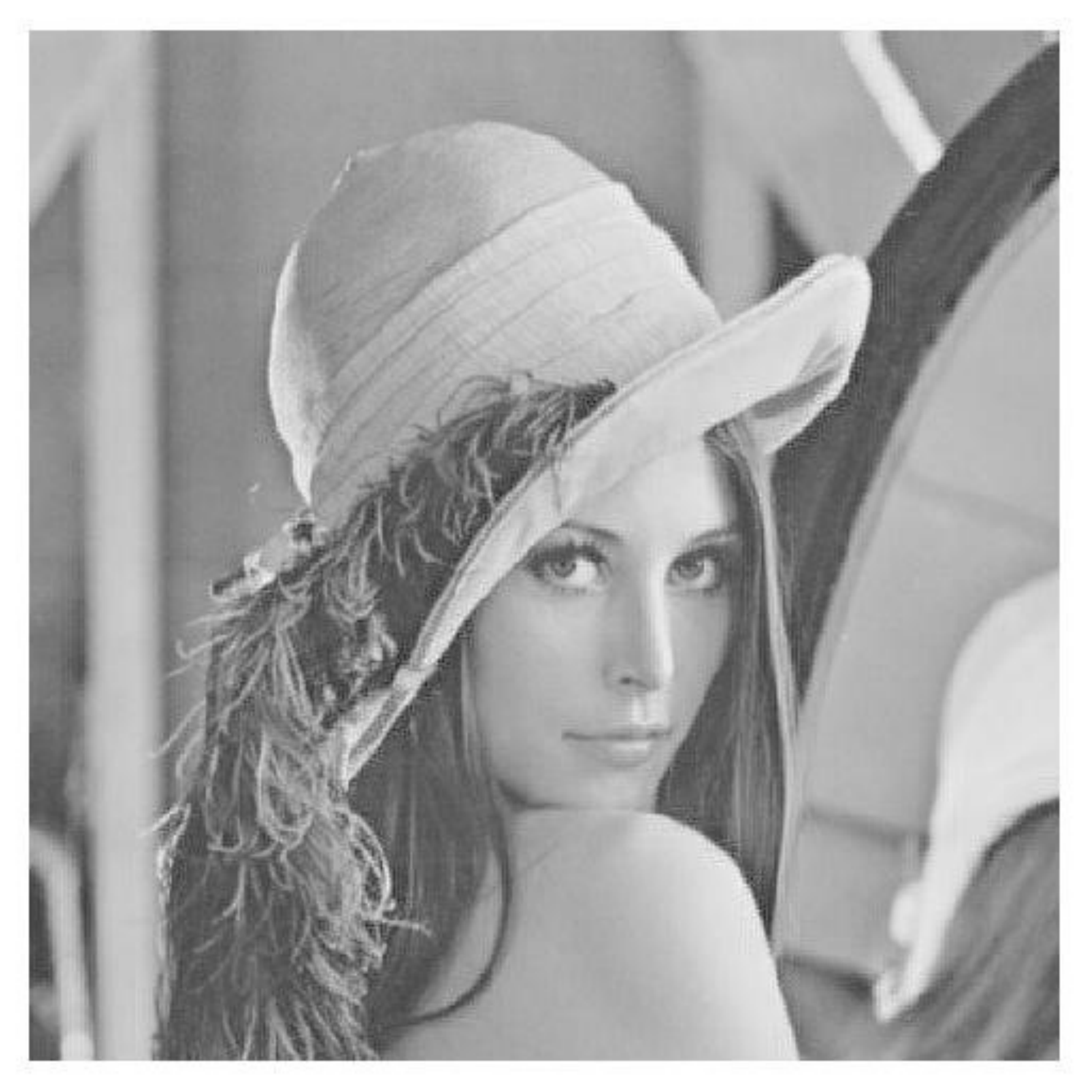}}
\subfigure[BAS-2010]{\includegraphics[width=0.24\linewidth]{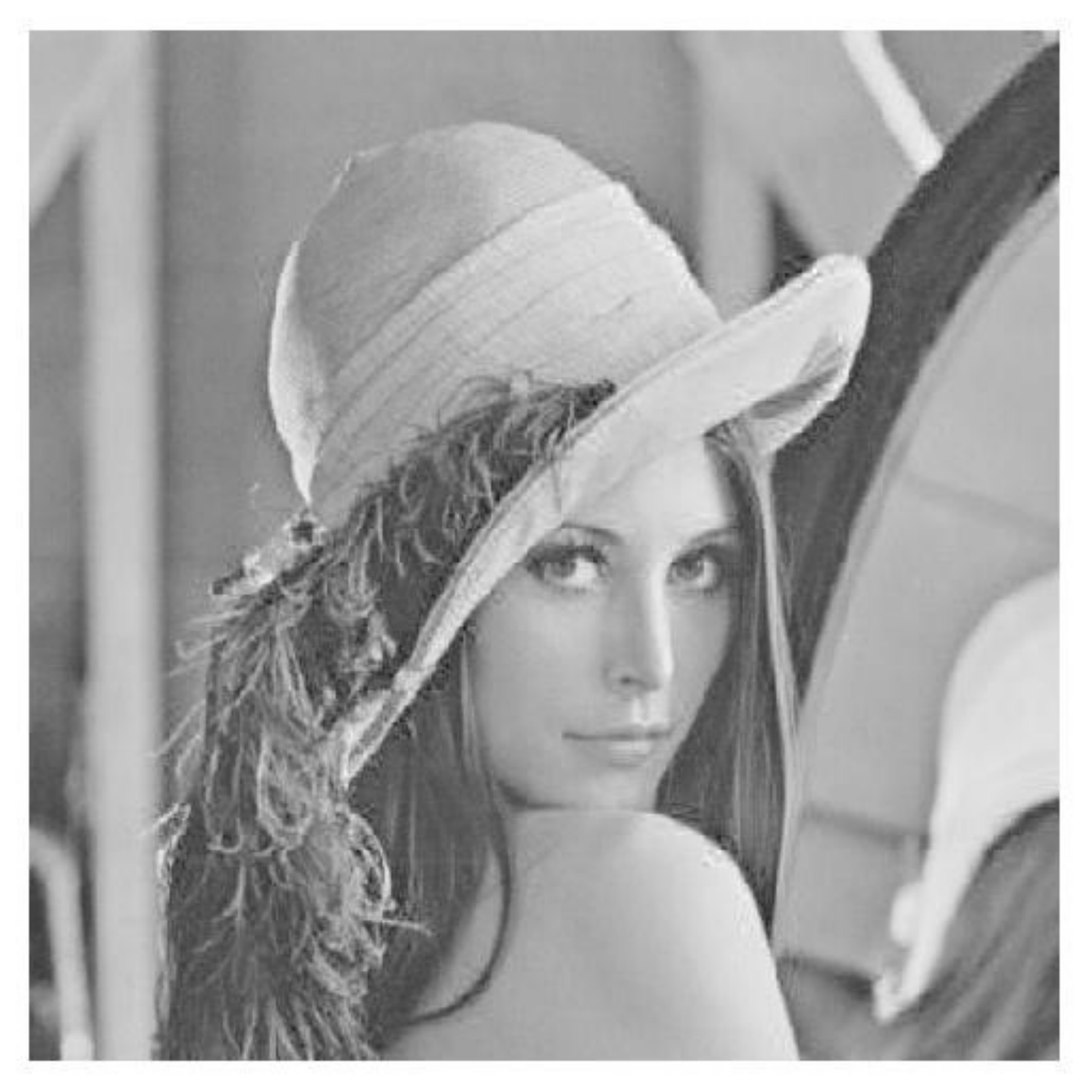}}
\subfigure[WHT]{\includegraphics[width=0.24\linewidth]{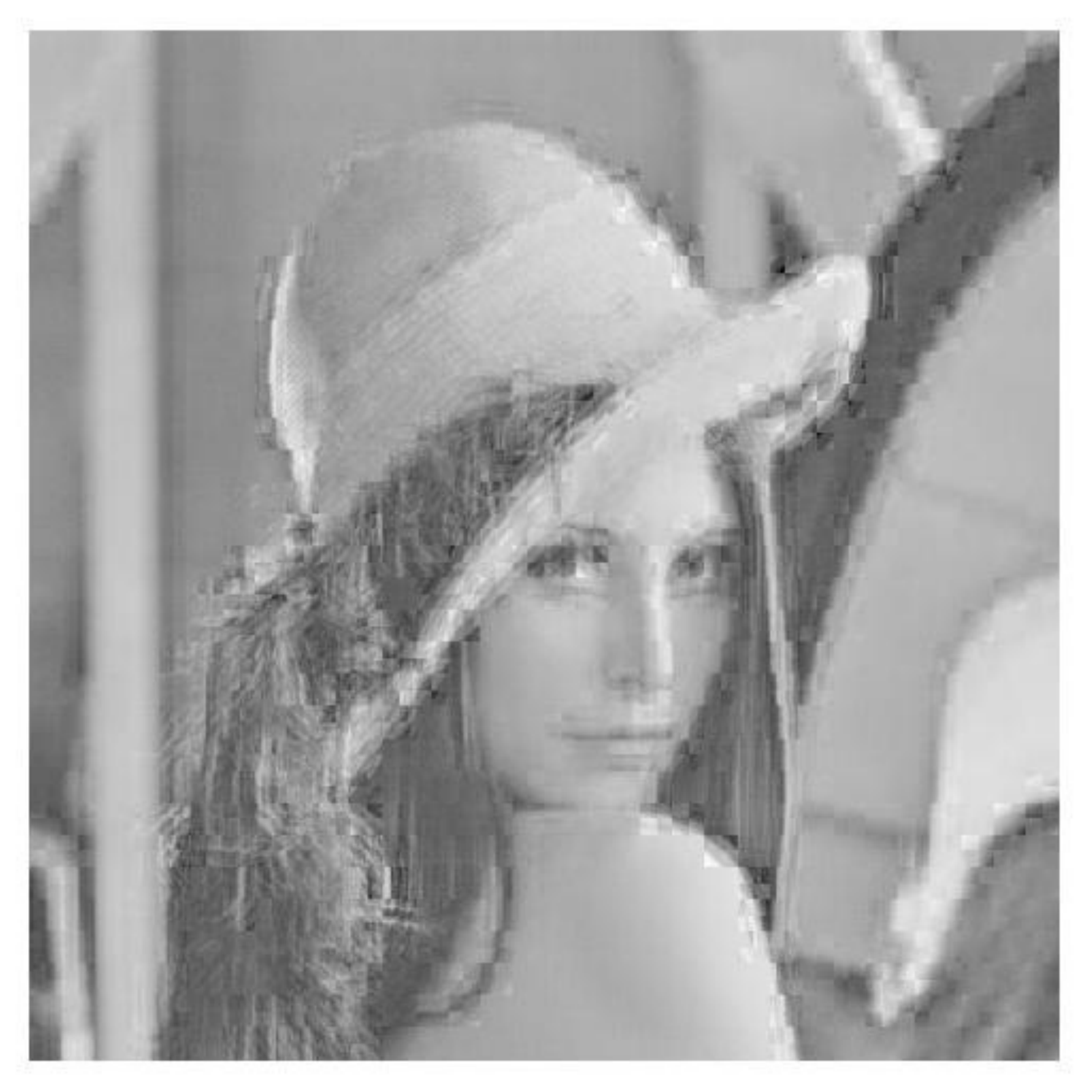}}

\subfigure[DCT]{\includegraphics[width=0.24\linewidth]{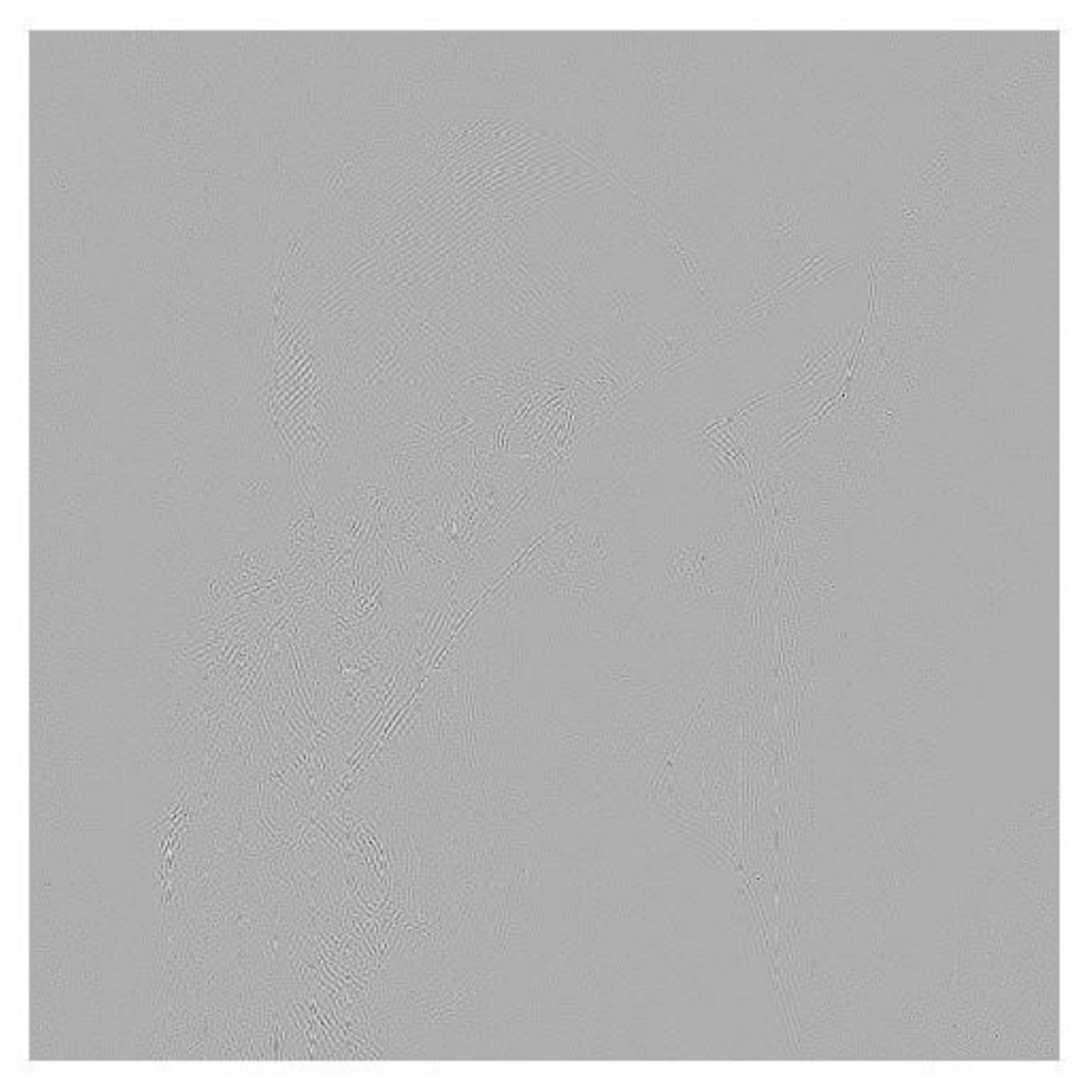}}
\subfigure[proposed]{\includegraphics[width=0.24\linewidth]{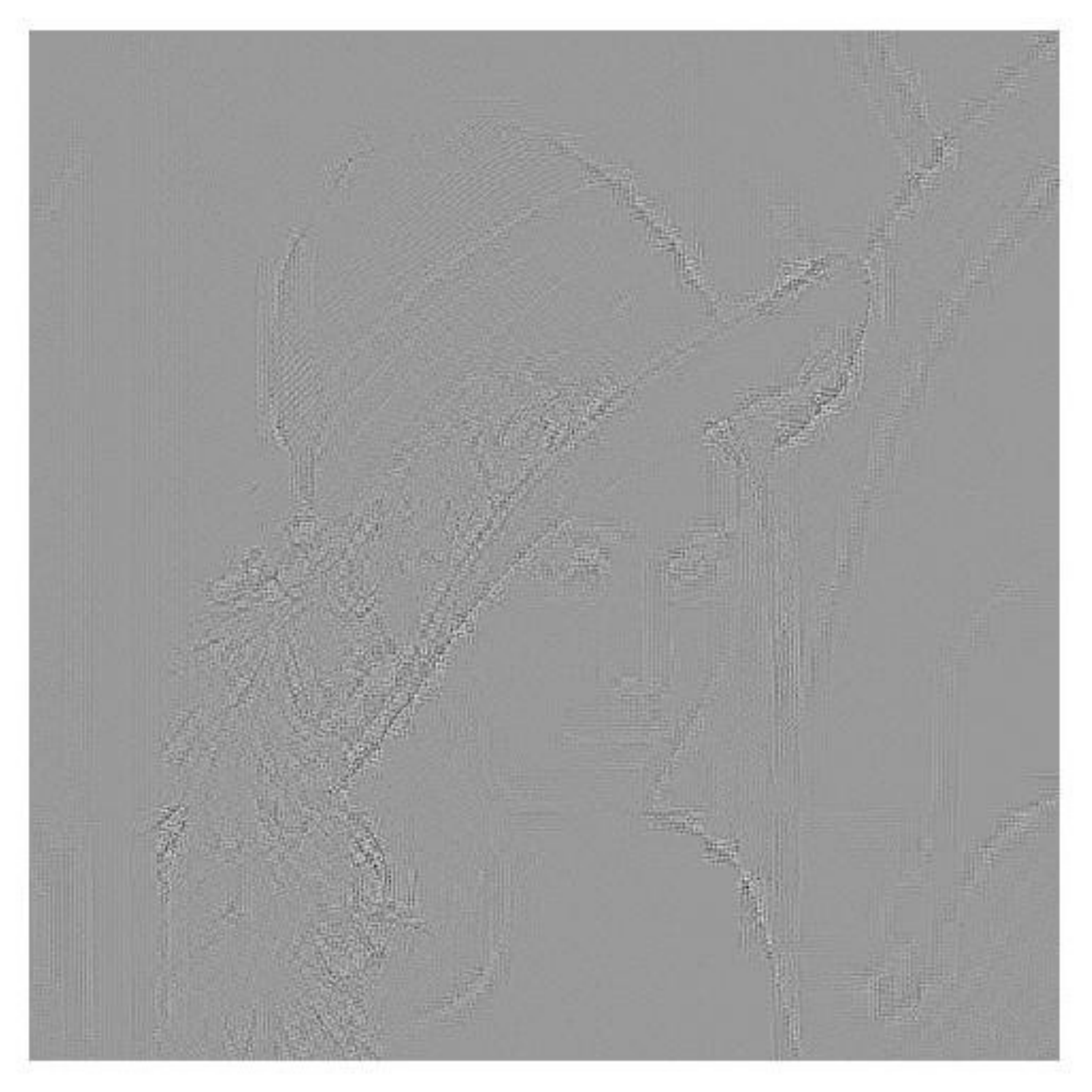}}
\subfigure[BAS-2010]{\includegraphics[width=0.24\linewidth]{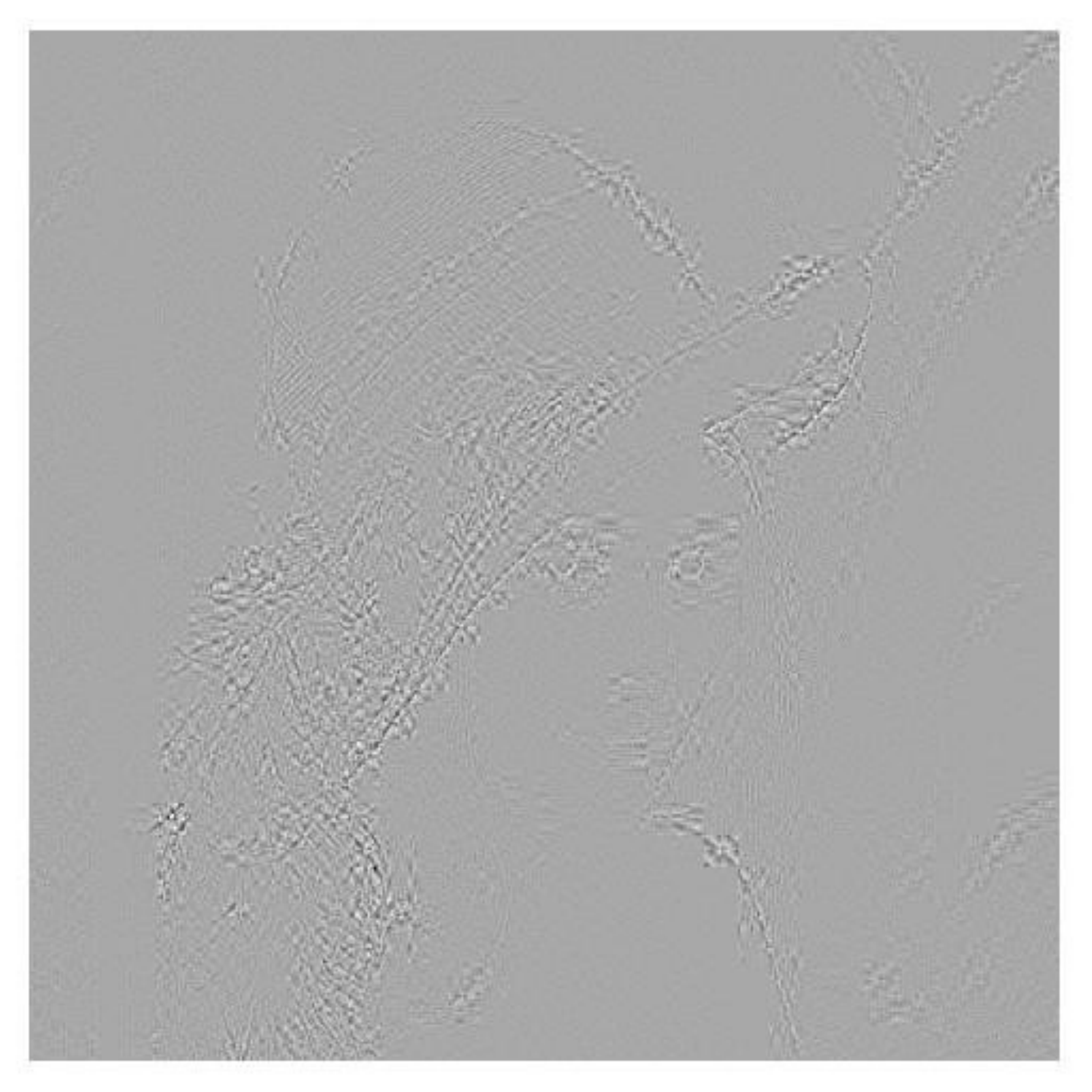}}
\subfigure[WHT]{\includegraphics[width=0.24\linewidth]{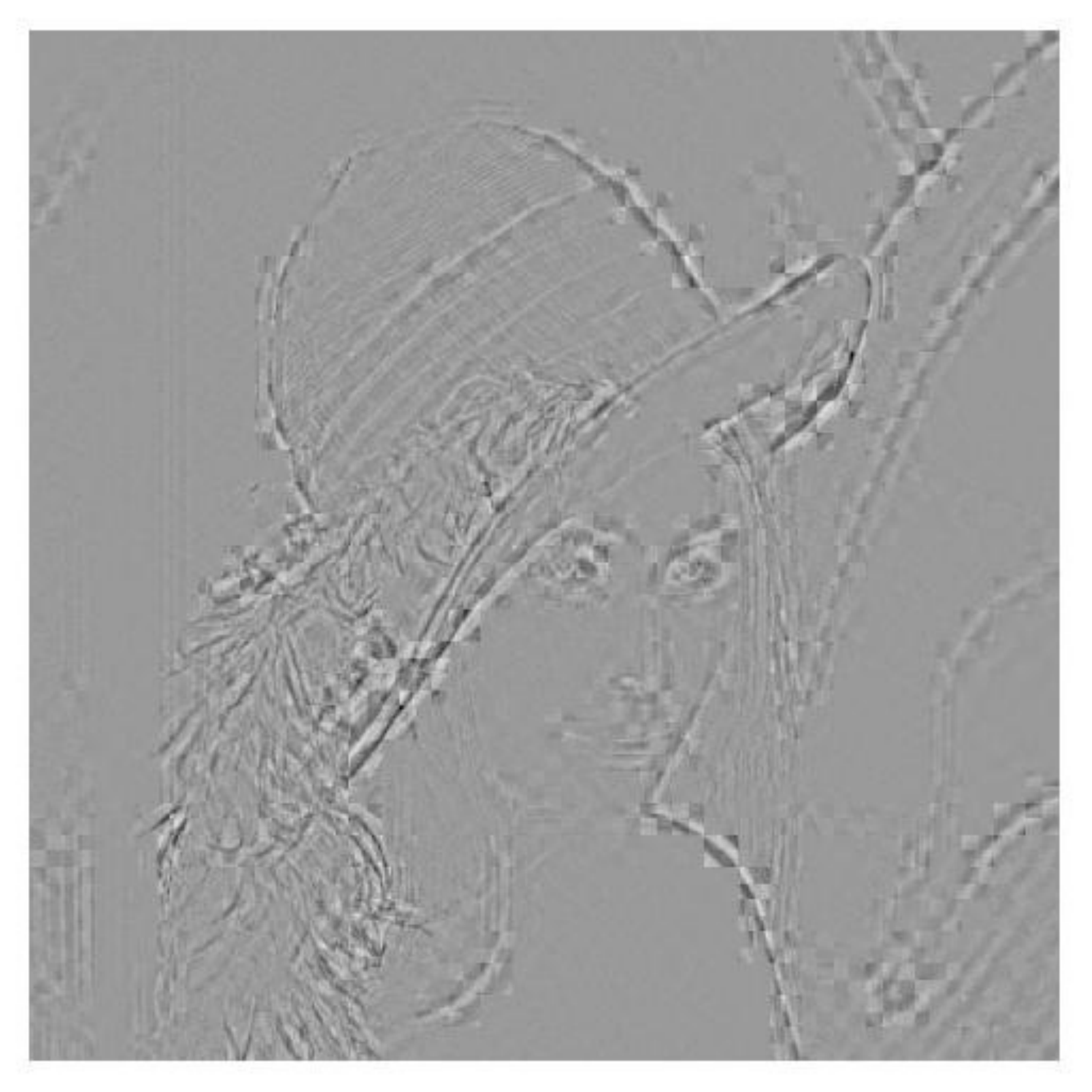}}

\caption{
Compressed images (a--d) and difference images (e--h)
using
the DCT,
the proposed transform,
the BAS-2010 approximation,
and the WHT for the Lena image,
considering $r=100$.}
\label{f:lena}
\end{figure}

\begin{figure}
\centering

\subfigure[DCT]{\includegraphics[width=0.24\linewidth]{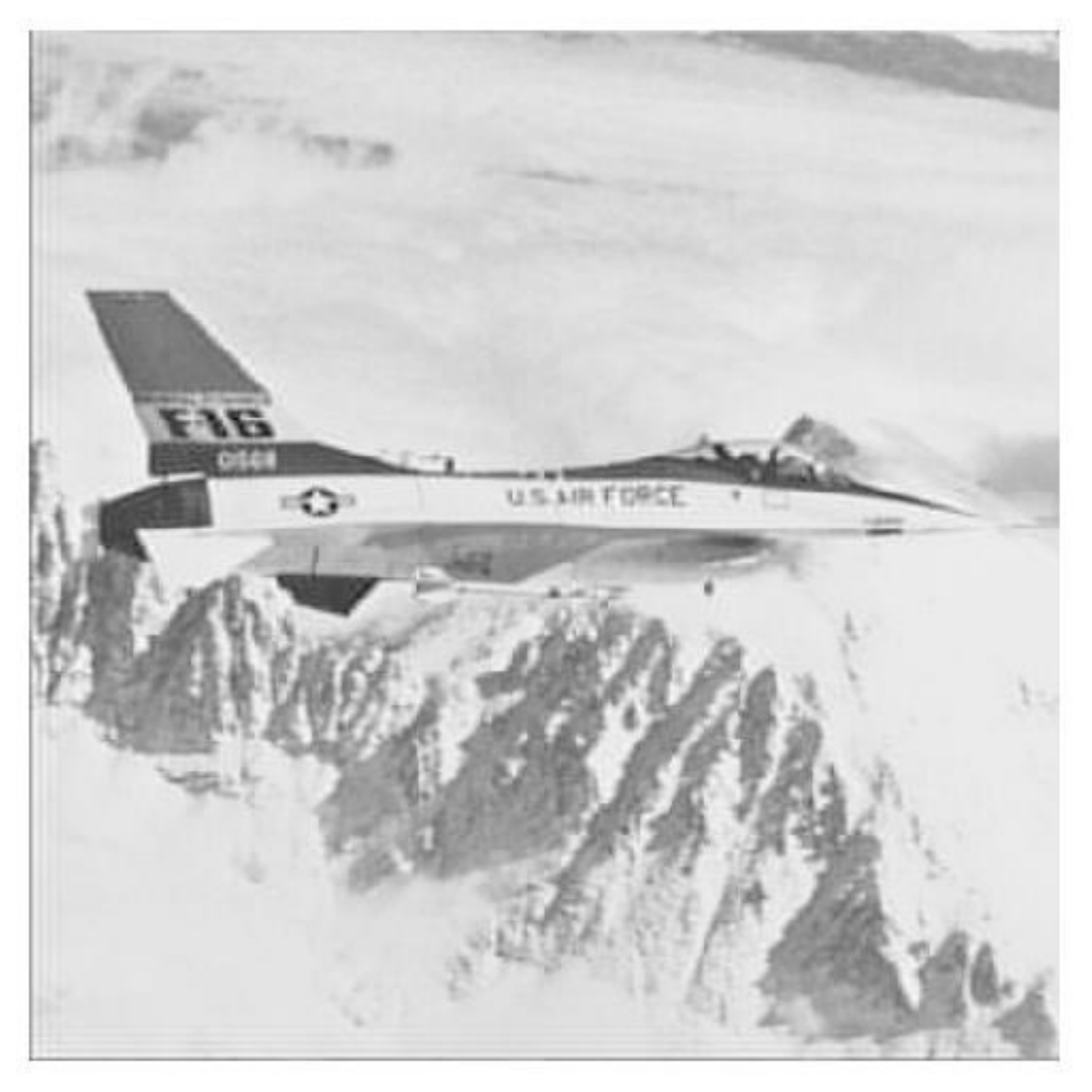}}
\subfigure[proposed]{\includegraphics[width=0.24\linewidth]{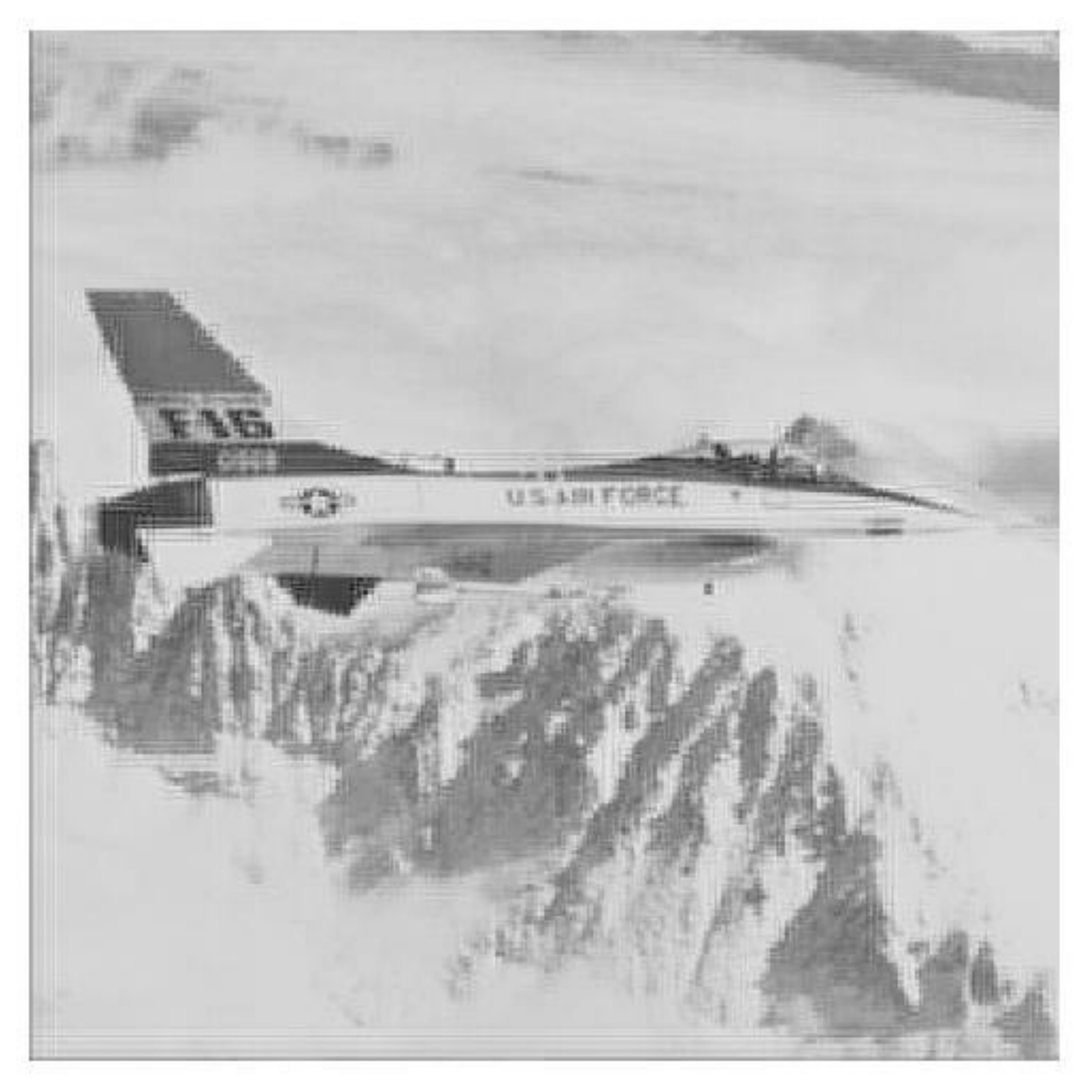}}
\subfigure[BAS-2010]{\includegraphics[width=0.24\linewidth]{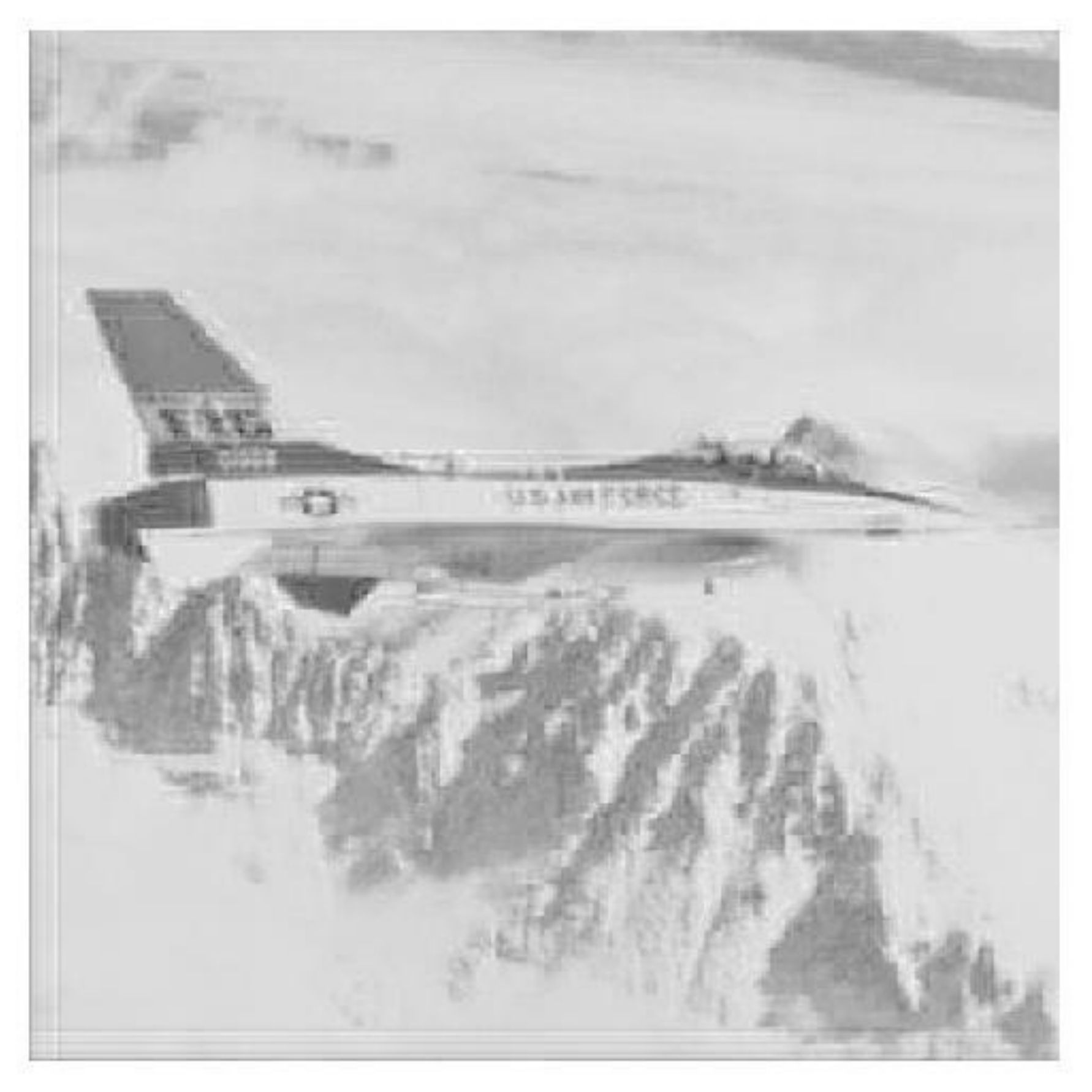}}
\subfigure[WHT]{\includegraphics[width=0.24\linewidth]{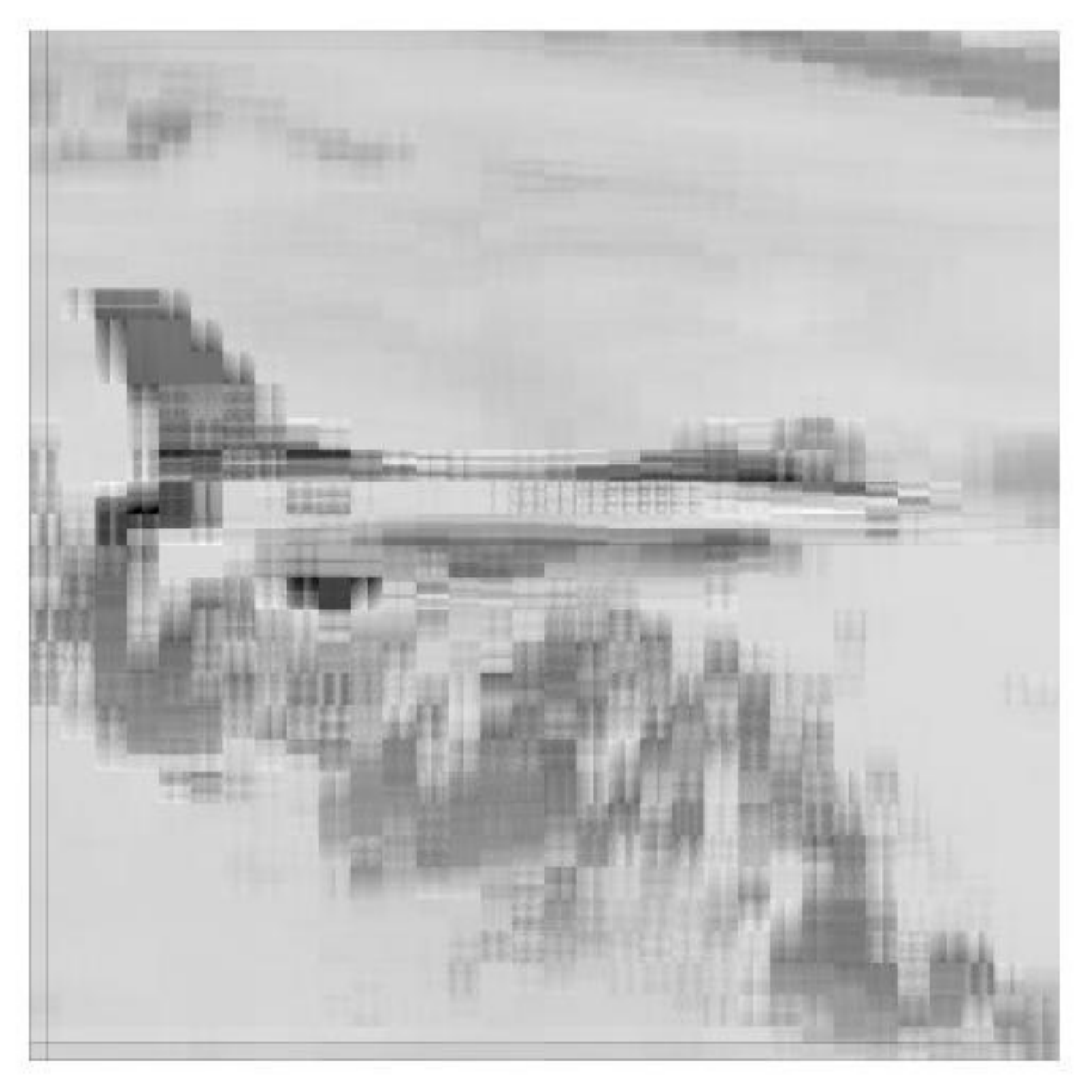}}

\subfigure[DCT]{\includegraphics[width=0.24\linewidth]{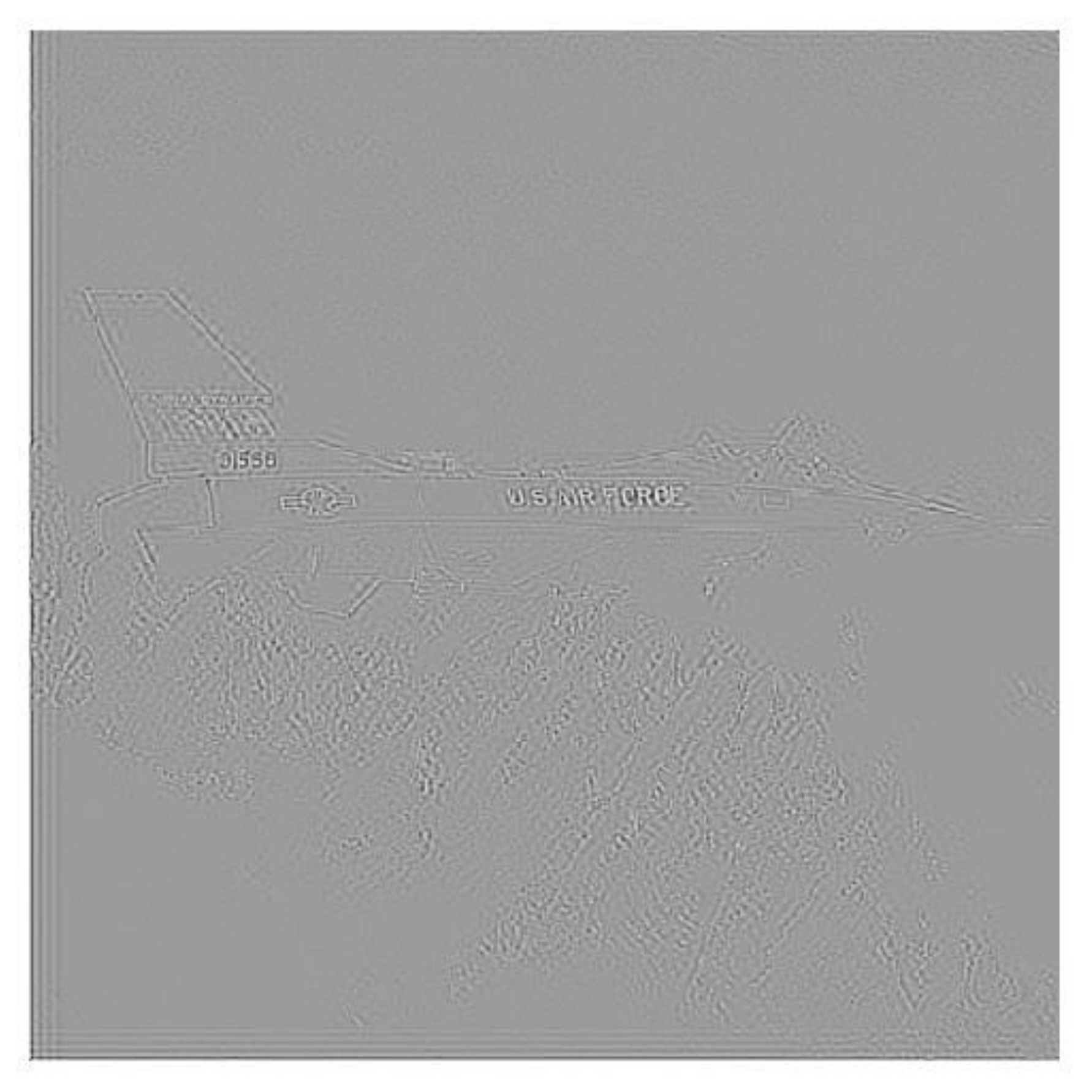}}
\subfigure[proposed]{\includegraphics[width=0.24\linewidth]{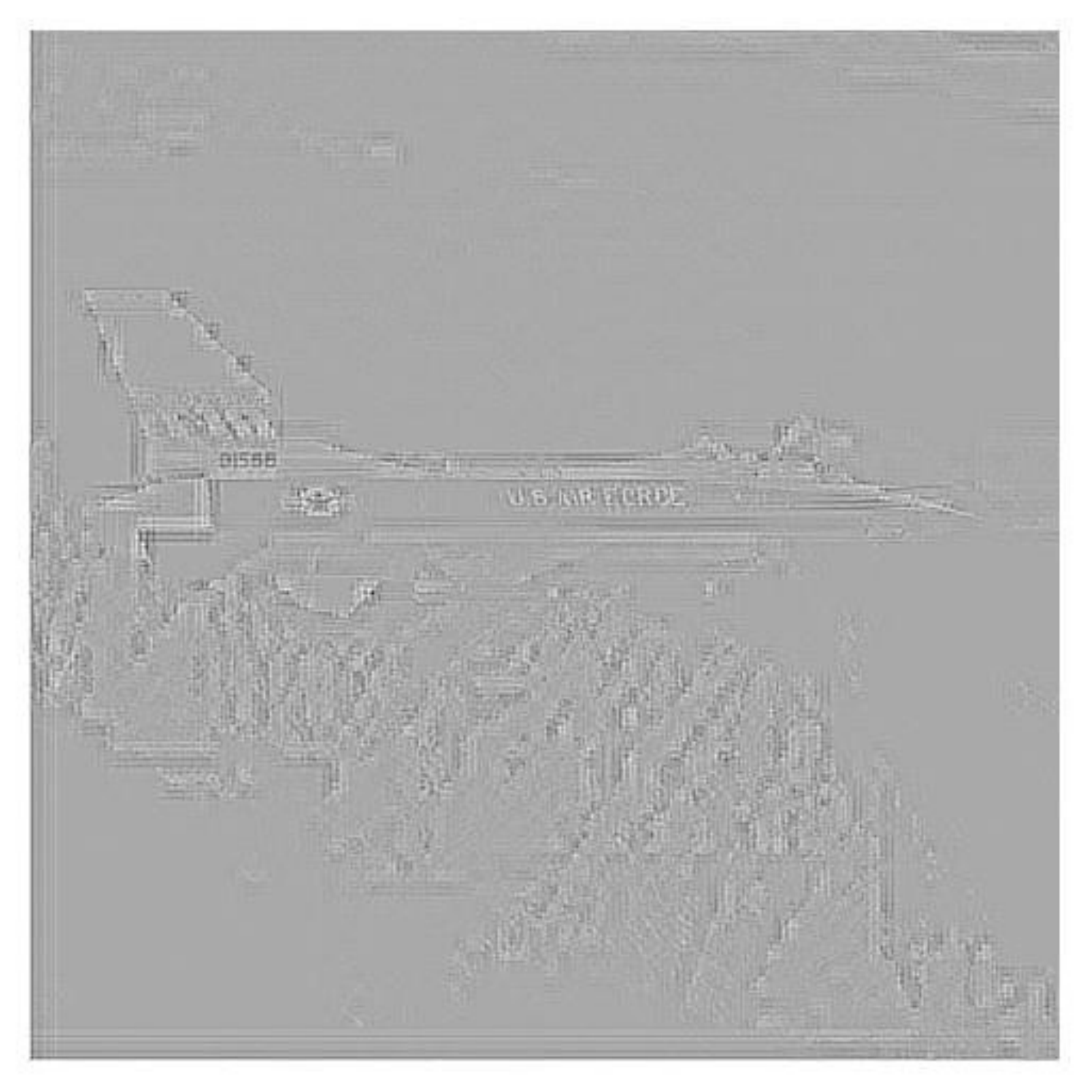}}
\subfigure[BAS-2010]{\includegraphics[width=0.24\linewidth]{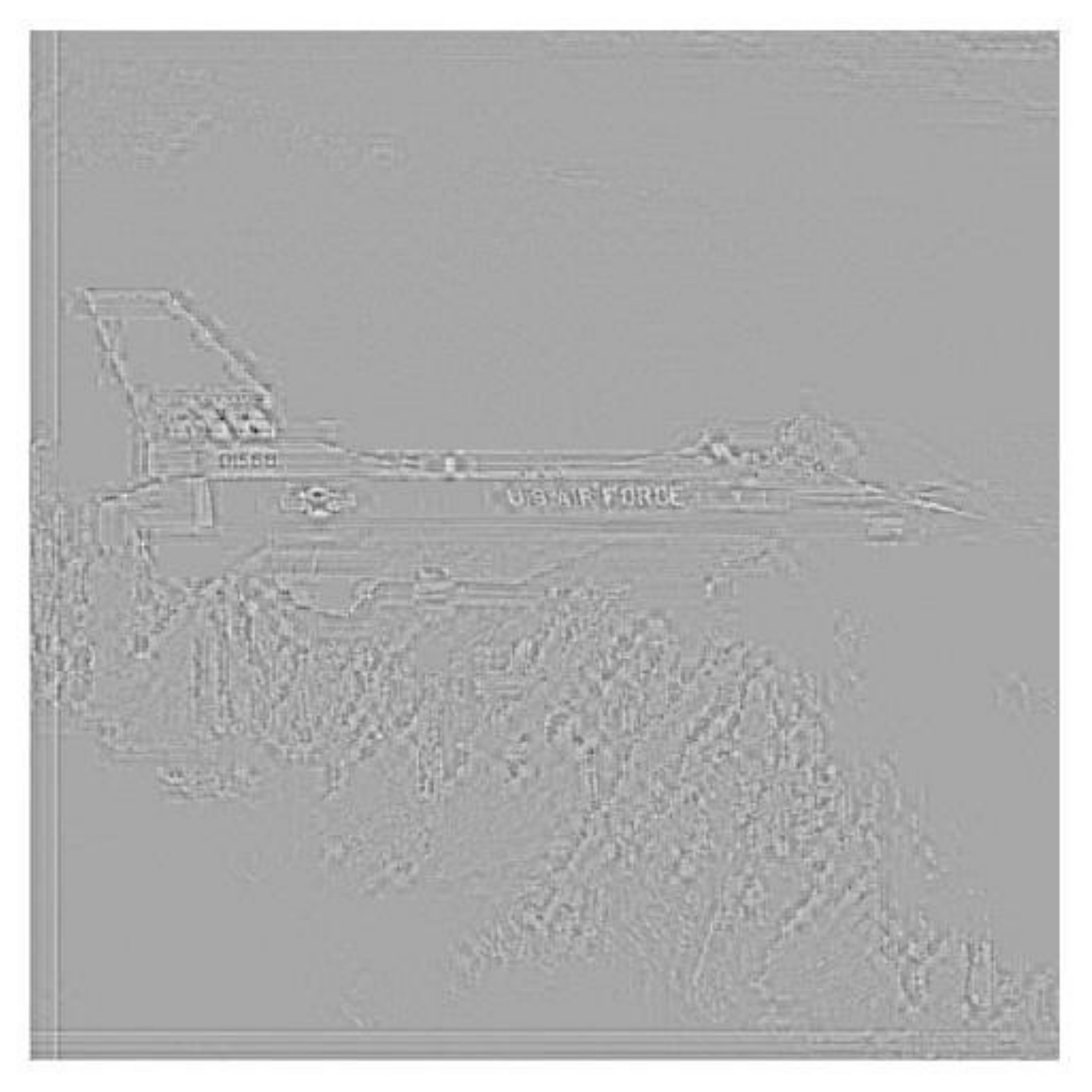}}
\subfigure[WHT]{\includegraphics[width=0.24\linewidth]{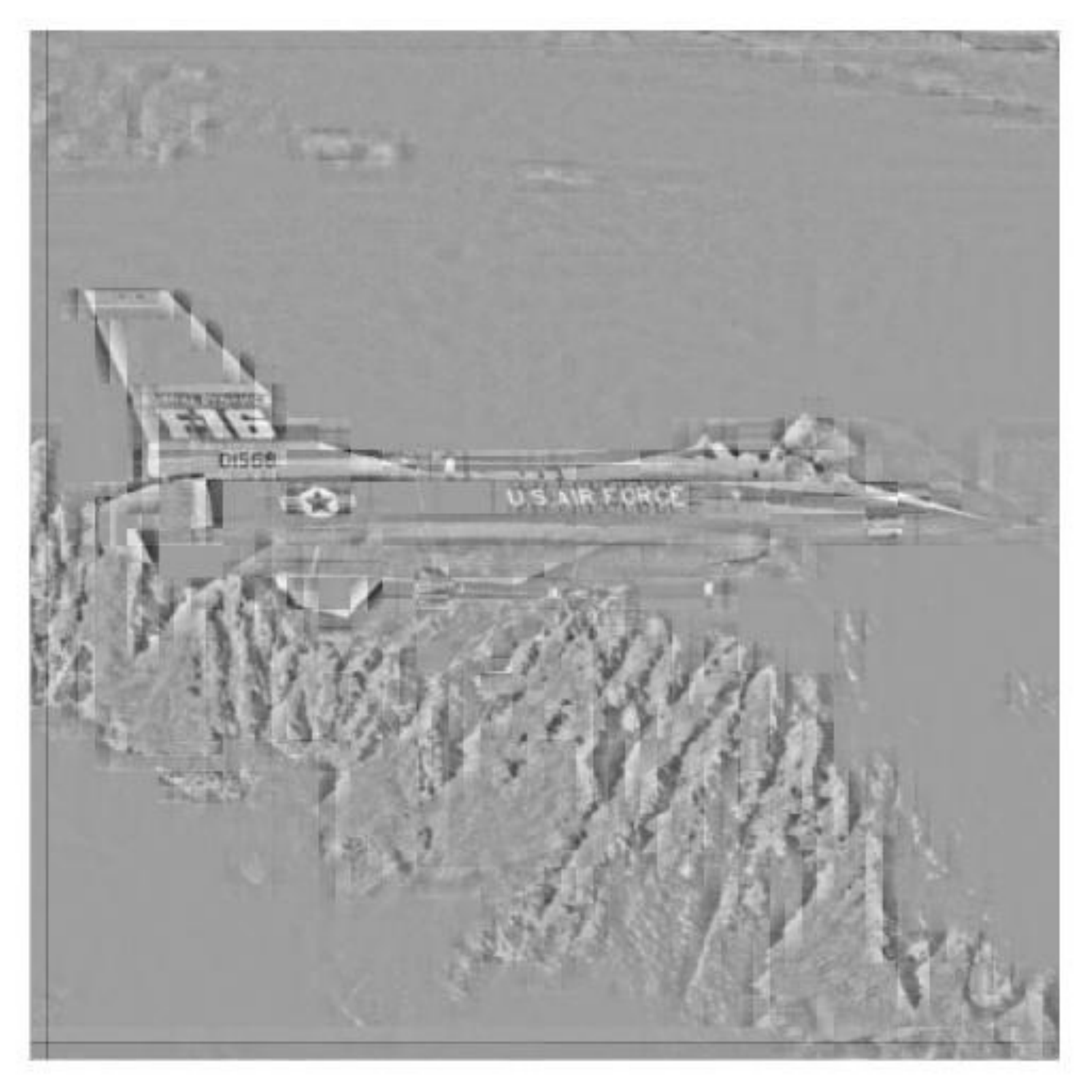}}

\caption{
Compressed images (a--d) and difference images (e--h)
using
the DCT,
the proposed transform,
the BAS-2010 approximation,
and the WHT for the Airplane (F-16)
image,
considering $r=40$.}
\label{f:f16}
\end{figure}

In order to convey a qualitative analysis,
Figures \ref{f:lena} and \ref{f:f16}
show
two standard images compressed according to the considered transforms.
The associate differences with respect to the original
uncompressed images are also displayed.
For better visualization,
difference images were scaled by a factor of two.
This procedure is routine and described in further detail in~\cite[p.~273]{salomon2004data}.
The images compressed with
the proposed transform are visually more similar
to the images compressed with DCT than the others.
As expected,
the WHT exhibits a poor performance.

\section{FPGA-based Hardware Implementation}
\label{section-implementation}

In this section,
the proposed DCT approximation and
the BAS-2010 algorithm~\cite{bas2010}
were physically implemented on a field programmable gate array~(FPGA) device.
We employed the 40~nm CMOS Xilinx Virtex-6 XC6VLX240T FPGA
for algorithm evaluation and comparison.
Beforehand
it is expected that the proposed algorithm exhibit
modestly
higher hardware demands.
This is due to the fact that
it
requires 72~additions,
whereas the BAS-2010 algorithms demands 64~additions.

We furnished circuit performance using metrics of
(i)~area~($A$) based on
the quantity of required
elementary programmable logic blocks (slices),
the number of look-up tables (LUTs),
and
the flip-flops count,
(ii)~the speed, using the critical path delay~($T$),
and
(iii)~the dynamic power consumption.

The number of occupied slices
furnished an estimate of the on-chip silicon real estate requirement,
whereas number
of LUTs and flip-flops are the main logic resources available in a slice.
In Xilinx FPGAs,
a LUT is employed as a combinational function generator
that can implement a given boolean function and
a flip-flop is utilized as a 1-bit register.
The critical path delay corresponds to
the delay associated with the longest combinational path
and
directly governs the operating frequency of the hardware.
The total power consumption of the hardware design constitutes of static and dynamic components.
Static power consumption in FPGAs is dominated by the leakage power of the logic fabric and the configuration static RAM.
Thus, is mostly design independent.
On the other hand,
the dynamic power consumption,
which accounts for the dynamic power dissipation associated with
clocks, logic blocks, and signals,
provides a metric for the power efficiency of
a given design~\cite{embedded_MM}.
The respective results are shown in
Tables~\ref{area_results},~\ref{speed_results}, and \ref{power_results},
where the metrics corresponding to each design
were measured for several choices of finite precision
using input word length $W \in \{4, 8, 12, 16\}$.

From Table~\ref{area_results},
it is observed that
the proposed design consumes
$\approx 10\%$ more LUTs hardware resources than~\cite{bas2010}.
For $W = 8$ the proposed design shows a $\approx 20\%$ increase in the number of slices consumed (and 10\% more LUTs) while gaining \mbox{1-2}~dB of improvement in PSNR compared to~\cite{bas2010}.
The increase in area shown by the proposed design has led to an increase in
the critical path delay,
area-time ($AT$),
area-time-squared ($AT^2$) metrics
and
to a higher power consumption as indicated
in Tables~\ref{speed_results} and~\ref{power_results}.
Of particular interest is the case for $W = 8$ input word size,
where the proposed algorithm and hardware design shows a $5.7\%$ and $10\%$ increase in the critical path delay and dynamic power consumption, respectively, when compared to the algorithm in~\cite{bas2010}.

In this paper,
we define a metric
consisting
of the product of
error figures and the $AT$ value:
\begin{equation*}
(\mbox{error figure}) \times (\mbox{area-time product})
.
\end{equation*}
The considered error figure
can be the $1/\mbox{PSNR}$, MSE, $1/\mbox{UQI}$,
or the total error energy,
as given in Table~\ref{t:error}.
This metric aims at combining
both the mathematical and the hardware
aspects of the resulting implementation.
The total error energy has the advantage of being
image independent,
being adopted in the combined metric.
Considering
the proposed architecture
and~\cite{bas2010},
the obtained values for the combined metric
are shown in Table~\ref{cost_metric}.

Although the proposed DCT approximation
consumes more resources
than~\cite{bas2010},
a much better approximation for the exact DCT is achieved
(see Table~\ref{t:error}).
This leads to superior compressed image quality (see Fig.~\ref{f:quality}).
Indeed,
the choice of algorithm is always a compromise between
its mathematical properties,
such as DCT proximity, energy error, and resulting image quality;
and
the related hardware aspects,
such as
area,
speed,
and
power consumption.
This implies our proposed algorithm
is a better choice over~\cite{bas2010}
when picture quality is of higher importance.
\begin{table*}
\centering
\caption{Area utilization for FPGA implementation.
}
\label{area_results}
\begin{tabular}{ccccccc}
\toprule
Input &
\multicolumn{6}{c}{Area}\\
\cmidrule{2-7}
word &
\multicolumn{3}{c}{BAS-2010~\cite{bas2010}} &
\multicolumn{3}{c}{Proposed}\\
\cmidrule{2-7}
length &
Registers & LUTs & Slices & Registers & LUTs & Slices\\
\midrule
4 &
403  & 543  & 172  &
597  & 524  & 178
\\
\midrule
8 &
828  & 704  & 241  &
956  & 909  & 290
\\
\midrule
12 &
1128  &  958  & 317  &
1253  & 1316  & 384
\\
\midrule
16 &
1432  & 1243  & 390  &
1597  & 1676  & 491
\\
\toprule
\end{tabular}
\end{table*}

\begin{table*}
\centering
\caption{Speed, $AT$, and $AT^2$ metrics for FPGA implementation.}
\label{speed_results}
\begin{tabular}{ccccccc}
\toprule
Input word &
\multicolumn{2}{c}{Speed (MHz)} &
\multicolumn{2}{c}{$AT$ ($\mbox{Slices} \cdot \mu s)$} &
\multicolumn{2}{c}{$AT^2$ ($\mbox{Slices}\cdot{\mu s}^2\cdot 10^{-3}$)}
\\
\cmidrule{2-7}
length & BAS-2010~\cite{bas2010} & Proposed & BAS-2010~\cite{bas2010} & Proposed & BAS-2010~\cite{bas2010} & Proposed\\
\midrule
4 & 369.13 & 342.9 & 0.466 & 0.519 & 1.26 & 1.51\\
\midrule
8 & 361.92 & 342.114 & 0.666 & 0.848 & 1.84 & 2.48\\
\midrule
12 & 363.63 & 336.813 & 0.872 & 1.140 & 2.40 & 3.38\\
\midrule
16 & 341.29 & 338.18 & 1.143 & 1.452 & 3.35 & 4.29\\
\toprule
\end{tabular}
\end{table*}

\begin{table*}
\centering
\caption{Dynamic power consumption for FPGA implementation.}
\label{power_results}
\begin{tabular}{ccccccccc}
\toprule
Input &
\multicolumn{8}{c}{Dynamic Power (mW)}\\
\cmidrule{2-9}
word &
\multicolumn{4}{c}{BAS-2010~\cite{bas2010}} &
\multicolumn{4}{c}{Proposed}
\\
\cmidrule{2-9}
length & Clocks & Logic & Signals & Total & Clocks & Logic & Signals & Total\\
\midrule
4 & 0.033 & 0.022 & 0.030 & 0.085 & 0.040 & 0.020 & 0.029 & 0.089\\
\midrule
8 & 0.041 & 0.016 & 0.034 & 0.091 & 0.041 & 0.023 & 0.037 & 0.101\\
\midrule
12 & 0.059 & 0.022 & 0.050 & 0.131 & 0.054 & 0.033 & 0.054 & 0.141\\
\midrule
16 & 0.069 & 0.028 & 0.070 & 0.167 & 0.065 & 0.042 & 0.077 & 0.184\\
\toprule
\end{tabular}
\end{table*}

\begin{table}
\centering
\caption{Comparison of the cost associated with each design}
\label{cost_metric}
\begin{tabular}{ccc}
\toprule
Input & \multicolumn{2}{c}{Combined metric}\\
\cmidrule{2-3}
word length & BAS-2010~\cite{bas2010} & Proposed\\
\midrule
4 & 31.22 & 4.19\\
\midrule
8 & 44.62 & 6.85\\
\midrule
12 & 58.42 & 9.21\\
\midrule
16 & 76.57 & 11.73\\
\toprule
\end{tabular}
\end{table}

\section{Conclusion}
\label{section-conclusion}

This paper introduced a new 16-point DCT approximation.
The proposed transform requires no multiplication
or bit shifting operations,
is orthogonal,
and
its matrix elements are only $\{-1, 0, 1\}$.
Using spectral analysis methods
described in~\cite{haweel2001new,cintra2011dct},
we demonstrated that the proposed transform
outperforms the WHT and the BAS-2010
as an approximation for the 16-point DCT.
The proposed transform was considered into
standard image compression methods.
The resulting images were assessed for quality by means of
PSNR, MSE, and UQI.
According to these
metrics,
the proposed transform
could outperform the WHT and the BAS-2010 approximation
at any compression ratio.
We also
derived an efficient fast algorithm
for the proposed matrix,
which required 72 additions.

This algorithm was implemented in hardware
and compared with
a state-of-the-art 16-point DCT approximation~\cite{bas2010}.
FPGA-based rapid prototypes were
designed,
simulated,
physically implemented,
and tested for 4-, 8-, 12-, and 16-bit input data word sizes.
A typical application having 8-bit input image data could be subject to
16-point DCT approximations
at a real-time rate of $342\cdot10^6$ transforms per second,
for FPGA clock frequency of 342~MHz,
leading to a pixel rate of $5.488\cdot 10^9$ pixels/second.
Both proposed and BAS-2010 algorithms
were realized on FPGA and tested and hardware metrics including
area,
power,
critical path delay,
and
area-time complexity.
Additionally,
an extensive investigation of relative performance
in both subjective mode as well as objective picture quality metrics
using
average PSNR,
average MSE,
and
average UQI
was produced.
The proposed DCT approximation algorithm
improves on the state-of-art algorithm in~\cite{bas2010}
by 1-2 dB for PSNR at the cost of only eight extra adders.

Video coding using motion partitions
larger than 8$\times$8 pixels is investigated
in~\cite{davies2010suggestion, telecommunicationstandardizationsector2009video}
with satisfactory application in
H.264/AVC standard for video compression.
In this perspective, the new proposed approximation transform
is a candidate technique to image and
video coding with block size equal to 16$\times$16.
This blocklength is of particular importance in the emerging H.265 reconfigurable video codec standard~\cite{kalva2006}.

\section*{Acknowledgments}

This work was partially supported by CNPq and FACEPE (Brazil);
and by the College of Engineering at the University of Akron, Akron, OH, USA.

{\small
\bibliographystyle{abbrv}
\bibliography{16-dct-app-clean}
}

\end{document}